\documentclass[prb,showpacs,aps,twocolumn]{revtex4}
\usepackage{amsmath}
\usepackage{graphicx}
\usepackage{dcolumn}
\usepackage{amssymb}
\usepackage{dcolumn}
\usepackage{color}

\begin{document}

\title{ Topology of quantum discord}
\author{Nga T. T. Nguyen}
\author{Robert Joynt}
\affiliation{Department of Physics, University of Wisconsin, Madison, Wisconsin 53706, USA}

\begin{abstract}
Quantum discord is an important measure of quantum correlations that can
serve as a resource for certain types of quantum information processing. \
Like entanglement, discord is subject to destruction by external noise. The
routes by which this destruction can take place depends on the shape of the
hypersurface of zero discord $\mathcal{C}$ in the space of generalized Bloch
vectors. For 2 qubits, we show that with a few points subtracted, this
hypersurface is a simply-connected 9-dimensional manifold embedded in a
15-dimensional background space. \ We do this by constructing an explicit\
homeomorphism from a known manifold to the subtracted version of $\mathcal{C}
$. \ \ We also construct a coordinate map on $\mathcal{C}$ that can be used
for integration or other purposes. This topological characterization of $%
\mathcal{C}$ has important implications for the classification of the
possible time evolutions of discord in physical models. \ The classification
for discord contrasts sharply with the possible evolutions of entanglement.
\ Using topological methods, we classify the possible joint evolutions of
entanglement and discord. \ There are 9 allowed categories: 6 categories for
a Markovian process and 3 categories for a non-Markovian process,
respectively. We illustrate these conclusions with an anisotropic XY spin
model. \ All 9 categories can be obtained by adjusting parameters. \ 
\end{abstract}

\maketitle

\section{Introduction}

\bigskip

Some of the most characteristic features of quantum mechanics show up in the
correlations of two subsystems that are independently measurable. The most
famous is entanglement\cite{Schrodinger}, but this notion does not exhaust
everything that is quantum about correlations. Even two systems that are
separable have zero entanglement can violate Bayes theorem, something that
cannot happen in classical physics. One quantity that measures the
additional quantumness of correlations is quantum discord $D$\cite%
{Vedral,Zurek}. Roughly speaking $D$ is the difference between the total
correlation once entanglement has been subtracted out, and the purely
classical correlation. Discord can serve as a resource for the
accomplishment of certain tasks in a way somewhat similar to the way that
entanglement can. For computation, the quantum algorithm DQC1 does seem to
use discord rather than entanglement\cite{Laflamme,Datta,Datta2}, and the
same is true for dense coding\cite{Bennett}. More general statements about
the uses of discord are difficult to make at this stage. Quantum discord and
other quantum correlation measures have recently received an extensive review%
\cite{Modi}.

\bigskip

One question that is of experimental importance is how quantum correlations
are erased by external noise. In the case of entanglement, there is a rather
rich range of possible behaviors of the time evolution of the concurrence
[the function $C(t);t\in \{0,\infty \}$] as a composite system loses its
quantum correlations\cite{Yu}. A somewhat similar, though distinctly more
limited, range of behaviors has been found in numerical studies of the
discord evolution [the function $D(t);t\in \{0,\infty \}$]\cite%
{Werlang_2009,Maniscalco,Ferraro2010,Fanchini,Fanchinivolume,Lang2011,Mazzola,Bellomo2012,Pal}%
. For entanglement, a general classification of time evolutions was seen to
depend on understanding the topology of entanglement: essentially the
structure of the set of separable states $\mathcal{S}$\cite{Zhou1,Zhou2}.
The purpose of the current work is to achieve the same goal for discord. We
will first determine the relevant topological properties of the set $%
\mathcal{C}$ of concordant states, i.e., the set of states for which $D$
vanishes, then deduce a general classification of the types of evolution of
the discord. Furthermore, we shall give examples of physical models that
realize the various types of evolution. \ The paper will focus on the case
of 2 qubits.

\bigskip

The most basic result about $\mathcal{C}$, established by Ferraro \textit{et
al.}\cite{Ferraro2010}, is that it is of zero 15-volume. \ To understand the
significance of this, we first note that the set of 2-qubit density
matrices, which we shall call $\mathcal{M}$, is a convex subset of a real
15-dimensional vector space. $\mathcal{M}$ itself is a 15-dimensional
manifold with boundary: any interior point of $\mathcal{M}$ has a
neighborhood that is homeomorphic to a neighborhood in $%
\mathbb{R}
^{15}$. $\mathcal{C}$ is a subset of $\mathcal{M}$. The fact that it has
zero 15-volume means that the dimension of any neighborhood of any point in $%
\mathcal{C}$ is less than 15, but gives no further information. We shall
show that (except for one point)\ the precise number for the local
dimensionality of $\mathcal{C}$ is 9. \ It has been shown previously that $%
\mathcal{C}$ is path-connected; we shall prove the stronger result that $%
\mathcal{C}$ (with one point removed) is simply connected. \ The zero-volume
statement already implies a very important point about discord evolution:
sudden death of discord is not possible. This was conjectured early on from
results of numerical studies and the connection with the geometry of $%
\mathcal{S}$ was understood. Other phenomena, such as frozen discord\cite%
{Maniscalco}, have also been shown to benefit from a geometric analysis\cite%
{Caves}. These analyses have been carried out in the 3-dimensional set of
Bell-diagonal states. 

Our aim here is to extend this framework to the full 15- dimensional space.
This will allow us to characterize in a topological fashion all joint
evolutions of entanglement and discord that lead to the disappearance of
both. Some evolutions have been computed by previous authors\cite%
{Roszak,Benedetti}.

The paper is organized as follows: \ Sec. II establishes concepts and
notation. \ Sec. III establishes the basic facts about the geometrical and
topological nature of $\mathcal{C}$. \ Sec. III applies the results of Sec.
II to the dynamical evolution of the discord, first establishing a
categorization of the possible evolutions, then illustrating this
categorization. \ In Sec. IV we give a discussion and the outlook for future
work.

\section{Discord, geometric discord, and frozen discord}

The definition of quantum discord that best expresses its foundation in
information theory is: 
\begin{equation*}
D(B|A)=I(A:B)-J(B|A),
\end{equation*}%
where $I(A:B)$ is the quantum mutual information:{} 
\begin{equation*}
I(A:B)=S(A)+S(B)-S(A,B),
\end{equation*}%
$S(A)$ is the usual von Neumann entropy, while $J(B|A)$ is a measure of the
total classical correlation present. $J(B|A)$ is defined in stages. First
note that if system $A$ is measured by an operator $E_{a}$ and is found to
be in the state $a$, then the density matrix of $B$ after the measurement is 
$\rho (B|a)=$Tr$_{A}(E_{a}\rho_{AB})/p_{a}$, where $p_{a}$ is the
probability of measuring the result $a$ in the state $\rho _{AB}$, i.e., $%
p_{a}=$Tr $(E_{a}\rho _{AB})$. We may then define a conditional entropy
under the measurement of $E_{a}:S(B|E_{a})=\sum_{a}p_{a}S(\rho _{B|a})$, and
then we have a corresponding mutual-information-like quantity $%
J(B|E_{a})=S(B)-S(B|E_{a})$. Quantum mechanics is distinguished from
classical mechanics by the fact that this quantity depends on the choice of
measurements. To remove this ambiguity, we maximize over the choice of $%
\{E_{a}\}$ and arrive at a measure of the total classical correlation $%
J(B|A)=$max$_{\{E_{a}\}}J(B|E_{a})$. $D(B|A)$ is clearly not symmetric
between systems $A$ and $B$, but it has the essential property of being
invariant under local unitary operations.

For our purposes, its most important property of discord is that $D(B|A)=0$
when $\rho _{AB}$ is classical-quantum: $\rho _{AB}=\sum_{a}p_a \Pi
_{a}\otimes \rho \left( B|a\right) $. Here $\{\Pi_a\}$ is any set of
rank-one projectors and $\rho \left( B|a\right) $ is the resulting partial
density matrix for $B$ if the result has been obtained from a measurement of 
$A$. This gives an explicit definition of the set $\mathcal{C} $ of
concordant states mentioned above.

We intend to investigate the topology of $\mathcal{C}.$ \ To define a
topology on any set requires a specification of its open subsets. \ A metric
is the most convenient way to do this, and we will employ the metric on the
set $\mathcal{M}$ of density matrices that follows from the Hilbert-Schmidt
inner product: \ $\left( \rho ,\rho ^{\prime }\right) =Tr(\rho \rho ^{\prime
})$. \ To give a consistent treatment of discord, we also need a
metric-based definition. \ Fortunately, there is the geometric discord,
defined by%
\begin{equation}  \label{eq:geo_discord}
D_{G}\left( B|A\right) =\min_{\chi \in \mathcal{C}}\left\vert \rho
_{AB}-\chi \right\vert =\min_{\chi \in \mathcal{C}}~\text{Tr}\left[ \left(
\rho _{AB}-\chi \right) ^{2}\right] ,
\end{equation}%
i.e., $D_{G}\left( \rho _{AB}\right) $ is the Hilbert-Schmidt distance from $%
\rho _{AB}$ to the nearest point of $\mathcal{C}$. This differs slightly
from the information-theory based definition above. \ We will comment on the
differences below. \ 
\begin{figure}[h]
\begin{center}
\vspace*{0.1cm} \includegraphics[width=7.9cm]{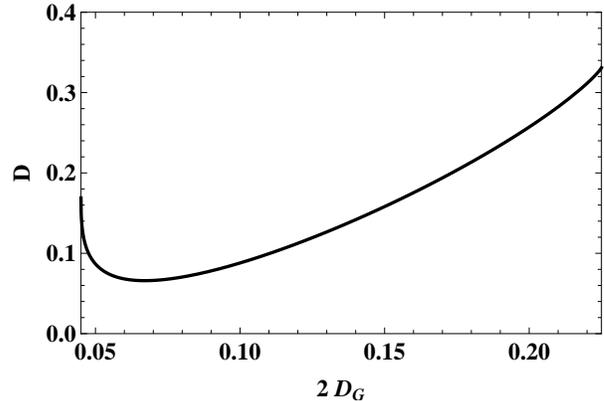}
\end{center}
\par
\vspace{-0.4cm}
\caption{The quantum discord is plotted as a function of normalized
geometric discord for a trajectory that lies in the space of Bell-diagonal
states. The trajectory $\ $has constant $N_{11}=-0.7$, $N_{22}=-0.3$ and $%
N_{33}$ is an implicit variable along the curve. \ This trajectory is shown
in Fig. 2. \ Note that a monotonically increasing $D_{G}$ does not imply an
increasing $D.$}
\label{fig:dis_geo-diss_nonmono}
\end{figure}
Since we intend to compare entanglement and discord, we need a corresponding
metrical definition of entanglement, the geometric entanglement: 
\begin{equation*}
E_{G}\left( B|A\right) =\min_{\chi \in \mathcal{S}}\left\vert \rho
_{AB}-\chi \right\vert =\min_{\chi \in \mathcal{S}}~\text{Tr}\left[ \left(
\rho _{AB}-\chi \right) ^{2}\right] ,
\end{equation*}%
where $\mathcal{S}$\ is the set of separable states, i.e. $\rho _{AB}\in 
\mathcal{S}$ if and only if 
\begin{equation*}
\rho _{AB}=\sum_{a}p_{a}\rho _{A}^{a}\otimes \rho _{B}^{a},
\end{equation*}%
where the $p_{a}$ are probabilities and $\rho _{A}^{a},\rho _{B}^{a}$ refer
to systems $A$ and $B$, respectively. \ We shall also have occasion to refer
to classical states, which we take to be states of the form 
\begin{equation*}
\rho _{AB}=\sum_{a}p_{a}\Pi _{A}^{a}\otimes \Pi _{B}^{a},
\end{equation*}%
where $\Pi _{A}^{a},\Pi _{B}^{a}$ are projections. \ The set of pure states,
for which there is a basis in which $\rho _{AB}$ is itself a projection
operator, will be denoted by $\mathcal{P}$. \ 

For 2 qubits, a general state can be written using the basis of SU(4)
generators: 
\begin{eqnarray}
\rho &=&\frac{1}{4}(\sigma _{0}\otimes \sigma _{0}+\sum_{i=1}^{3}{%
N_{0i}\sigma _{0}\otimes \sigma _{i}+\sum_{i=1}^{3}N_{i0}{\sigma _{i}}}%
\otimes \sigma _{0}  \notag  \label{eq:general_rho} \\
&&+\sum_{i,j=1}^{3}N_{ij}\sigma _{i}\otimes \sigma _{j}).
\end{eqnarray}
$\sigma _{0}$ is the $2\times 2$ identity and $\sigma _{1,2,3}$ are the
Pauli matrices that generate $SU(2)$. \ The 15 $SU(4)$ generators are $%
\sigma _{i}\otimes \sigma _{j}$ (where either $i>0$ or $j>0).$ \ $N_{0i}$
and $N_{i0}$ are sometimes called local Bloch vectors of qubit $A$ and $B,$
respectively. \ $N_{ij}$ with both $i>0$ and $j>0$ is sometimes termed the
correlation tensor. \ This representation of the density matrix is variously
called the Pauli basis, the polarization vector, the coherence vector, and
the generalized Bloch vector. \ We will usually use the latter term. 

Since we will mainly use the geometric discord in this paper, it is
important to clarify the distinction between \ the usual quantum discord and
the geometric discord. \ Unlike a quantum entanglement measure such as the
concurrence and its geometric counterpart (distance to the nearest separable
state), discord and geometric discord are not always monotonic functions of
one another, i.e., it is possible that $dD/dt$ has the opposite sign from $%
dD_{G}/dt$ at points along some trajectory $\rho _{AB}\left( t\right) $ in
the state vector space. \ An example is shown in Fig.~\ref%
{fig:dis_geo-diss_nonmono} where quantum discord and geometric discord show
different behavior for a trajectory restricted to the Bell-diagonal subclass
of states defined by the fact that only the three components $%
N_{11},N_{22},N_{33}$ are non-zero. \ In Fig. \ref{fig:dis_geo-diss_nonmono}%
, the trajectory moves along the straight line $N_{11}=-0.7,$ $%
N_{22}=-0.3,N_{33}=-1+2t$ as $t$ varies from 0 to 1. \ It can be seen that
there are values of $t$ such that $dD/dt<0$ but $dD_{G}\left( t\right) >0.$
\ 
\begin{figure}[htb]
\begin{center}
\vspace{0.2cm} \includegraphics[width=8.5cm]{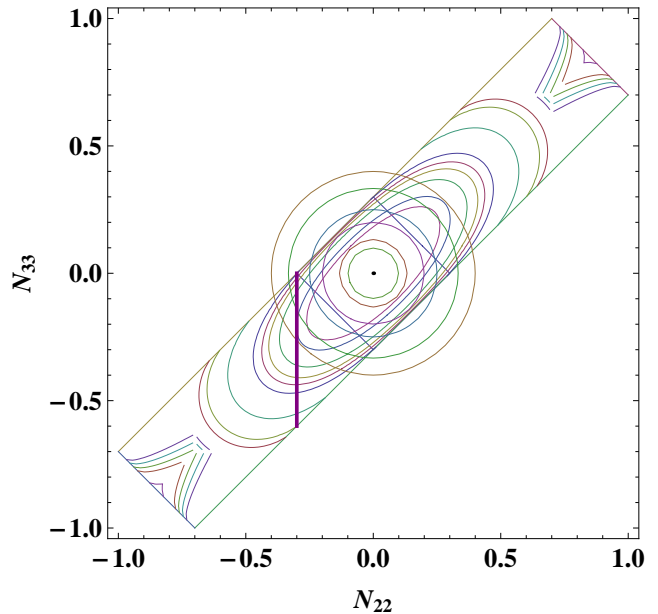}
\end{center}
\par
\vspace{-0.3cm}
\caption{(Color online) The plane of Bell-digonal states having $%
N_{11}=-0.7. $ Circles centered on the origin represent surfaces of constant
geometric discord. \ Other more complex curves represent surfaces of
constant discord. \ Only the states lying inside the tilted rectangle are
physical states that satisfy positivity. \ Geometry of the Bell-diagonal
subclass of states having characterized as the titled rectangle. The square
is the corresponding separable subset of this subclass of states. \ The
larger the (geometric) discord value, the further the constant (geometric)
discord curve from the concordant (zero-discord) point ($N_{11}, N_{22},
N_{33}$) =(-0.7, 0, 0). The vertical line with the segment inside the
rectangle describes one possible trajectory that results in discord and its
geometric measure of the system not mutually monotonic increasing with one
another. This is the trajectory shown in Fig.~\protect\ref%
{fig:dis_geo-diss_nonmono} }
\label{fig:geo-diss_nonmono}
\end{figure}

The reason for this non-intuitive behavior can be seen from Fig.~\ref%
{fig:geo-diss_nonmono}, where curves of constant $D\ $and $D_{G}$ in the
plane defined by $N_{11}=-0.7$ are depicted. \ All allowed states then lie
inside the tilted rectangle in this plane. \ The only concordant point in
this plane is $(N_{11},N_{22},N_{33})=\left( -0.7,0,0\right) $ - the center
of the tilted rectangle. \ The curves of constant geometric discord are the
circles. \ The other more complicated curves are the curves of constant
quantum discord. \ The trajectory of Fig.~\ref{fig:dis_geo-diss_nonmono} is
the thick vertical line segment $N_{22}=-0.3,$ \ staying inside the
rectangle of the vertical line plotted in Fig.~\ref{fig:geo-diss_nonmono}. \
This trajectory hits some of the geometric discord curves only once while it
hits some of the discord curves two times, which is the reason for the two
different time behaviors. \ It is easily seen that the trajectory must be
carefully chosen for this to occur, which is the reason for the
arbitrary-seeming values of the trajectory parameters. 
One can see from this discussion that while the two quantities $D$ and $%
D_{G} $ measure essentially the same thing, subtle differences in the actual
functional dependences mean that the relation between the two is not
monotone. 
\begin{figure}[h]
\begin{center}
\vspace*{-0.5cm} \hspace{-1.41cm} \includegraphics[width=10.0cm]{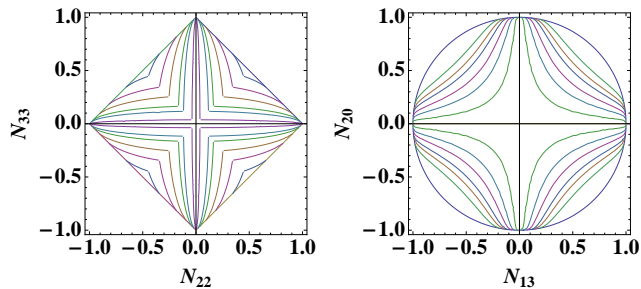}
\end{center}
\par
\vspace{-3.5cm}
\caption{(Color online) Examples of curves of constant discord for two
different sections of $N$-space. \ Only the two coordinates listed are
nonzero. \ Coordinate axes are always straight surfaces of zero discord, and
discord increases as the distance from the axes increases, but the precise
functional dependence varies depending on which axis pair is considered. \
All states inside the square and the disk are separable. }
\label{fig:frozenUdiscordZZ}
\end{figure}
\newline
\textbf{Frozen discord} \newline
\textquotedblleft Frozen" quantum discord occurs when $D\left( t\right) $ or 
$D_{G}\left( t\right) $ is constant positive number for a finite interval of
time. \ During this time period, the quantum mutual information and the
classical correlations decrease, but the difference $D=I-J_{\text{class}}$
remains fixed \cite{Maniscalco,Lim}. \ Since surfaces of zero discord can
have simple shapes in $N$-space\cite{Caves} surfaces of constant geometric
discord can also have relatively simple shapes and simple plausible models
can produce the phenomenon of frozen discord. \ This is much less likely to
occur for the quantum discord, for which the shapes of the surfaces are
typically complex. \ Examples of the latter are shown in Fig.~\ref%
{fig:frozenUdiscordZZ}.

\section{State space}

\subsection{Topology of $\mathcal{C}$}

\label{subsec:Top_C}

Optimizing the classical correlations requires considerable effort: closed
formulas for quantum discord have been obtained only for a few classes of
quantum states, typically the \textbf{X}-type class (see e.g. Refs.\cite%
{Luo,Ali,Girolami2011}). \ The geometric discord $D_{G}$, defined in Eq.~(%
\ref{eq:geo_discord}), is usually easier to compute. \ The minimization
present in definition (\ref{eq:geo_discord}) can now be performed explicitly
and the geometric discord is obtained in a fully analytical form\cite%
{Dakic,Girolami} 
\begin{equation}
D_{G}=\frac{1}{4}\left( \sum_{i=1}^{3}\sum_{\alpha =0}^{3}N_{i\alpha
}^{2}-k_{\text{max}}\right) ,  \label{eq:DG_analytical_form}
\end{equation}%
where $k_{\text{max}}$ is the maximum eigenvalue of the matrix%
\begin{equation}
L_{ij}=N_{i0}\left( N_{j0}\right) ^{T}+\sum_{k=1}^{3}N_{ik}N_{jk}.  \notag
\end{equation}%
\ \ We also note that the geometric discord satisfies\cite{Girolami,LuoFu} $%
1\geq 2D_{G}\geq D^{2}$ with equality corresponding to pure states of
maximally entangled. \ 

The density matrix of the zero-discord state for a pair of qubits A and B
has the form (details in appendix \ref{a:obtainmu}): 
\begin{equation}
\rho _{AB}= p \left\vert \Psi _{0}\right\rangle \left\langle \Psi _{0}\right
\vert \otimes \rho _{0} + (1-p) \left\vert \Psi _{1}\right\rangle
\left\langle \Psi _{1}\right \vert \otimes \rho _{1}.  \label{eq:rhoc}
\end{equation}%
$\rho _{k}$, $k=0, 1$, is a marginal density matrix for qubit $B.$ $\
D\left( \rho _{AB}\right) =D_{G}\left( \rho _{AB}\right) =0$ if and only if $%
\rho _{AB}$ has this classical-quantum form. \ If so, then $\rho _{AB}\in 
\mathcal{C},$ the set of concordant states. \ 

Our goal in this section is to determine topological structure of $\mathcal{C%
}$. \ We shall show that if certain points are subtracted from $\mathcal{C}$
we get a set $\mathcal{C}_{-}$ that is a boundaryless 9-manifold. \ Thus
nearly every point of $\mathcal{C}$ has a neighborhood that is homeomorphic
to an open set of $H^{9},$ the 9-dimensional half-space. \ This serves as a
basis for understanding the dynamics of discord. \ \ 

The strategy of the argument is first to establish a one-to-one continuous
and invertible mapping $f$ from a known boundaryless 9-manifold $\mathcal{A}$
to a set $\mathcal{C}_{-}$. \ We then consider extensions of $f$ in order to
understand the relation of $\mathcal{C}_{-}$ to $\mathcal{C}$ itself$.$ \ We
can also show that the 9 tangent vectors of this mapping are linearly
independent on $\mathcal{C}_{-}$ so that we have a valid coordinate chart on 
$\mathcal{C}_{-}.$ \ Since the difference between $\mathcal{C}$\ and $%
\mathcal{C}_{-}$ is a set of measure zero, the coordinate chart is
sufficient for purposes of, for example, integration on $\mathcal{C}$. \ 

We consider the set $\mathcal{A}=J\times S_{2}\times B_{3}\times B_{3}.$ \ $%
\times $ denotes the Cartesian product. $\ J,\ $a boundaryless 1-manifold,
is the open interval $\left( 0,1/2\right) .$ \ Points belonging to $J$ will
be labeled by $p$: $0<p<1/2.$ \ $S_{2},$ a boundaryless 2-manifold, is the
2-sphere. \ Points belonging to $S_{2}$ will be denoted by $\vec{m}$ or $%
\left( m_{1},m_{2},m_{3}\right) $ with $\left\vert \vec{m}\right\vert
^{2}=m_{1}^{2}+m_{2}^{2}+m_{3}^{2}=1.$ \ (Spherical polar coordinates will
also be used later ). $\ B_{3}$ is the \textit{open} 3-ball which is a
boundaryless 3-manifold. \ Points belonging to the first copy of $B_{3}$
will be denoted by $\vec{n}_{0}$ or $\left( n_{01},n_{02},n_{03}\right) $
with $\left\vert \vec{n}_{0}\right\vert
^{2}=n_{01}^{2}+n_{02}^{2}+n_{03}^{2}<1.$ \ Similarly for the second copy of 
$B_{3}$ and $\vec{n}_{1}.$ \ Since the Cartesian product of simply-connected
boundaryless manifolds is a boundaryless manifold, and the dimensions add, $%
\mathcal{A}$ is a simply-connected boundaryless 9-manifold. \ \ We now
define a map $f\left( p,\vec{m},\vec{n}_{0},\vec{n}_{1}\right) $ from $%
\mathcal{A}$ to $%
\mathbb{R}
^{15}$(Euclidean 15-space) $f:\mathcal{A\rightarrow }%
\mathbb{R}
^{15}$ by 
\begin{equation}
N_{0i}=pn_{0i}+\left( 1-p\right) n_{1i}  \label{eq:mu1}
\end{equation}

\begin{equation}
N_{i0}=\left( 2p-1\right) m_{i}  \label{eq:mu2}
\end{equation}

\begin{equation}
N_{ij}=m_{i}\left[ pn_{0j}-\left( 1-p\right) n_{1j}\right] .  \label{eq:mu3}
\end{equation}

The various $N$'s give the 15 components (appendix \ref{a:obtainmu}) of $f$
and $i,j=1,2,3.$ \ These can be thought of as a generalized Bloch vector for
states in $\mathcal{C}.$ \ It contains 3 components for the marginal density
matrices of the two individual qubits and 9 for the correlations. \
Geometrically, the points of $N_{i0}$, considered as a set in $%
\mathbb{R}
^{3},$ lie on the line joining the 3-vectors $\vec{n}_{0}$ and $\vec{n}_{1}$%
. \ Since $\ \vec{n}_{0}\in B_{3}$ and $\vec{n}_{1}\in B_{3},$ the set of
points $N_{i0}$ (i.e., the image of $f$ restricted to the first three
dimensions of $%
\mathbb{R}
^{15})$ fills out an open 3-ball $B_{3},$ and this set is \textit{%
independent of the value of} $p.$ \ Similarly the set of possible values of $%
N_{ij}$ for any fixed $i$ is an open 3-ball of radius $m_{i}$ that is
independent of $p.$ \ \ 

The physical meaning of the various parameters is clarified by computing the
magnitude of $\vec{N}:$ 
\begin{equation}
\left\vert \vec{N}\right\vert ^{2}=\left( 2p-1\right) ^{2}+2p^{2}\left\vert 
\vec{n}_{0}\right\vert ^{2}+2\left( 1-p\right) ^{2}\left\vert \vec{n}%
_{1}\right\vert ^{2}.  \label{eq:N_square}
\end{equation}%
Pure states have $\left\vert \vec{N}\right\vert ^{2}=3$ in our
normalization, which implies that the pure states of $\mathcal{C}$ have $p=0$
and $\left\vert \vec{n}_{1}\right\vert =1.$ \ Since entanglement and discord
are the same for pure states, these are product states, as is evident if we
insert the conditions for $p$ and $\vec{n}_{1}$ in Eq.~(\ref{eq:rhoc}) 

$f$ consists only of polynomial functions so it is obviously smooth. \ $%
\mathcal{C}_{-}$ is the image of $f$ and it is defined by Eqs.~(\ref{eq:mu1}%
), ~(\ref{eq:mu2}), ~(\ref{eq:mu3}), and the restrictions on the input
variables. \ im $f\subset 
\mathbb{R}
^{15}$ and $f$ is surjective on \ $\mathcal{C}_{-}$ by definition. \ \ $%
\mathcal{C}_{-}$ is clearly compact.\ 

It remains to show that $f$ is injective and therefore invertible. \ We note
first from Eq.~(\ref{eq:mu2})\ that $N_{i0},$ considered as a 3-vector $\vec{%
N}_{0}$, lies inside a ball of radius $1$: $%
N_{10}^{2}+N_{20}^{2}+N_{30}^{2}<1.$ \ This follows from the fact that $%
0<1-2p<1.$ It is also the case that any point in $%
\mathbb{R}
^{15}$ that has$~\left\{ N_{10},N_{20},N_{30}\right\} =\left\{ 0,0,0\right\} 
$ is not included in \ $\mathcal{C}_{-}$ since $\left\vert \vec{m}%
\right\vert =1$ and $p<1/2.$ \ We will comment on this later. \ The
restricted function $N_{i0}\left( p,m_{i}\right) $ is one-to-one for all\ $%
\left\{ N_{10},N_{20},N_{30}\right\} $ such that $%
0<N_{10}^{2}+N_{20}^{2}+N_{30}^{2}<1,$ and the inverse function is $\left(
m_{1},m_{2},m_{3}\right) =\left( N_{10},N_{20},N_{30}\right) /\left\vert 
\vec{N}_{0}\right\vert $ and $p=1/2-\left\vert \vec{N}_{0}\right\vert
/2\left\vert \vec{m}\right\vert .$ \ Hence the specification of $\vec{N}_{0}$
uniquely determines $p$ and $\vec{m}.$ \ Once these quantities are known and 
$N_{0i}$ and $N_{ij}$ are given, we can form the combinations 
\begin{eqnarray*}
\frac{1}{p}\left( N_{0i}+N_{ij}/m_{j}\right) &=&n_{0i} \\
\frac{1}{1-p}\left( N_{0i}-N_{ij}/m_{j}\right) &=&n_{1i}
\end{eqnarray*}%
obtained by adding and subtracting Eqs.~(\ref{eq:mu1}) and ~(\ref{eq:mu3}).
\ Because of the product form of $N_{ij},$ any choice of $j$ for which $%
m_{j}\neq 0$ (and at least one such must exist since $\left\vert \vec{m}%
\right\vert =1$) will do in these equations, which determine $n_{0i}$ and $%
n_{1i}$ uniquely. \ This completes the specification of $f^{-1}$. \ $f^{-1}$
maps every point in \ $\mathcal{C}_{-}$\ to a unique point of $\mathcal{A}.$
\ $f$ \ is injective and $f$ and $f^{-1}$ are continuous, so $f$ is a
homeomorphism. \ Every compact subset of $\mathcal{A}$ is mapped to a
compact subset of im $f,$ so $f$ is an embedding and \ $\mathcal{C}_{-}$ is
a boundaryless 9-manifold. \ Every point in $\mathcal{C}_{-}$ has a
neighborhood that is homeomorphic to a neighborhood in $%
\mathbb{R}
^{9}.$ \ 

The topology of im $f$ is found by a parallel argument. \ \ $\mathcal{C}_{-} 
$ is homeomorphic to $\mathcal{A}$, which is simply-connected since it is a
Cartesian product of simply-connected manifolds. \ Hence \ $\mathcal{C}_{-}$
is simply connected. \ Its algebraic topology is not entirely trivial,
however, since the second homology group $H_{2}\left( \mathcal{A}\right) =%
\mathbb{Z}
$ (because of the factor of $S_{2}$ in $\mathcal{A}),$ which implies that $%
H_{2}\left( \mathcal{C}_{-}\right) =%
\mathbb{Z}
$ as well. \ \ 

It remains to relate $\mathcal{C}_{-}$to $\mathcal{C},$ the set of
concordant states. \ To do so, we examine points in the closure of $\mathcal{%
A}$ and the associated extensions of $f.$ \ There are 3 classes of such
points, which we now consider in turn.

1. $\left\vert \vec{n}_{0}\right\vert =1$ and $\left\vert \vec{n}%
_{1}\right\vert =1.$ \ Addition of these points to $\mathcal{A}$ adds the
boundary of $B_{3}$ to the set of allowed $N_{i0}$ and similarly for the set
of allowed $N_{ji}$ whenever $m_{j}=1.$ \ The points added to $\mathcal{C}%
_{-}$ have neighborhoods homeomorphic to a neighborhood of a boundary point
of $H^{9},$ the 9-dimensional half-space, so they are typical boundary
points. \ Physically, $\left\vert \vec{n}_{0}\right\vert =1$ or $\left\vert 
\vec{n}_{1}\right\vert =1$ indicates a pure state of qubit $A$ in one term
of superposition. \ \ 

2. $p=0.$ For any continuous extension of $f$ \ to the points with $p=0$ we
find that the new points for the generalized Bloch vector are given by 
\begin{eqnarray*}
N_{0i} &=&n_{1i} \\
N_{i0} &=&-m_{i} \\
N_{ij} &=&-m_{i}n_{1j}.
\end{eqnarray*}%
Again, since the set of allowed $N_{0i}$ and $N_{ij}$ is independent of $p,$
the only effect of varying $p$ is to vary the magnitude of $N_{i0}$. \ $p=0$
corresponds to unit radius. \ Adding $p=0$ to the domain of $f$ thus adds
the boundary of $B_{3}$ to the set of allowed $N_{i0}$ and again these are
typical boundary points of \ $\mathcal{C}_{-}$.\ \ Physically, this value of 
$p$ corresponds to a product state: qubit $B,$ in a mixed state for all $%
p>0, $ is now in a pure state given by $\vec{m}$ and qubit $A$ is in an
arbitrary mixed state specified by $\vec{n}_{1}$. \ There is no correlation
whatever between $A$ and $B$. \ 

3. $\ p=1/2.$ \ These points also lie in the closure of $\mathcal{A}.$ Now
we obtain an extension of $f$ whose image includes the new points%
\begin{equation}
N_{0i}=\frac{1}{2}\left( n_{0i}+n_{1i}\right)
\end{equation}

\begin{equation}
N_{i0}=0
\end{equation}

\begin{equation}
N_{ij}=\frac{1}{2}m_{i}\left( n_{0j}-n_{1j}\right) .
\end{equation}%
\ We need only consider the change in the set of allowed $N_{i0},$ since the
set of allowed $N_{0i}$ and $N_{ij}$ is not affected by $p,$ as already
noted. \ The only points of $%
\mathbb{R}
^{15}$ that are added to im $f$ are those with $N_{i0}=0$ - otherwise there
is no change. \ For any fixed $p<1/2,$ \ $\mathcal{C}_{-}$ restricted to the
3-dimensional subspace $N_{0i}$ is an open 3-ball with the origin subtracted
out. For $p=1/2,$ the image of any extension of $f$ restricted to the
3-dimensional subspace $N_{i0}$ is the origin, for which $N_{0i},N_{i0}$ and
the $N_{ij}$ all vanish. \ The origin is a 0-dimensional object, so any
extension of $f$ that includes $p=1/2$ in its domain will not be invertible.
\ The origin does lie in $\mathcal{C},$ of course. \ However, it is easy to
show that it is not a simple boundary point. \ All of the 15 coordinate axes
belong to $\mathcal{C}$ and they intersect at the origin. \ This implies
that the origin does not have a neighborhood in $\mathcal{C}$ that is
homeomorphic to an open set of $%
\mathbb{R}
^{9}.$ \ Hence $\mathcal{C}$ itself is not a manifold. \ Physically, at $%
p=1/2$ the qubit $B$ is in the completely mixed state and any partial
density matrix is possible for qubit $A.$ \ \ \ \ 

The addition of points in classes 1 and 2 do not affect the algebraic
topology of $\mathcal{C}_{-}$. \ They are essentially boundary points and
any path passing through these points can be deformed into a path that lies
entirely in $\mathcal{C}_{-}$. \ This is probably also the case for points
in class 3, which leads to the conjecture that C itself is simply-connected.
\ We do not have a proof of this, however.

To summarize, we find that $\mathcal{C}_{-}\subset \mathcal{C}$ is a
simply-connected 9-manifold without boundary. \ The homeomorphism $f$
provides an embedding of $\mathcal{C}_{-}$ into the 15-dimensional space $%
\mathcal{M}$ of all density matrices, defining a 9-dimensional hypersurface
that differs from $\mathcal{C}$ itself by a set of meaure zero. \ 

Some additional properties of the hypersurfaces are:

(1) $\mathcal{C}_{-}$ includes points infinitesimally close to the origin,
(the point having $N_{i0}=N_{0i}=N_{ij}=0)$.\ \ $\mathcal{C}$ includes the
origin itself.

(2) $\mathcal{C}$ includes intervals lying on all 15 of the coordinate axes
(points for which only one of the $N_{i0},N_{0i},N_{ij}$ is nonzero). \ See
appendix \ref{a:15axes} for the proof. \ For example the $N_{0x}$ axis
corresponds to $p=1/2,n_{0x}=n_{1x}\neq 0,$ $n_{0y}=n_{0z}=n_{1y}=n_{1z}=0;$
the $N_{0x}$ axis corresponds to \ $m_{x}=1,m_{y}=m_{z}=0,$ $\vec{n}_{0}=%
\vec{n}_{1}=0.$ \ The $N_{xy}$ axis corresponds to $p=1/2,$\ $%
m_{x}=1,m_{y}=m_{z}=0,$ $n_{0y}=-n_{1y}\neq 0,$\ all others zero. \ 

(3) The four eigenvalues of $\rho _{AB}\in \mathcal{C}$ are:%
\begin{eqnarray*}
\lambda _{1, 2} &=&\frac{1}{2}p\left( 1\pm \left\vert \vec{n}_{0}\right\vert
^{2}\right) \\
\lambda _{3, 4} &=&\frac{1}{2}\left( 1-p\right) \left( 1\pm \left\vert \vec{n%
}_{1}\right\vert ^{2}\right).
\end{eqnarray*}%
Pure states have one eigenvalue equal to one and the others zero, which
means $p=0$, $\left\vert \vec{n}_{0}\right\vert =1$ and $\vec{n}_{1}=0.$\
These points lie on $\partial \mathcal{C}_{-},$ the boundary of $\mathcal{C}%
_{-}$. \ \ The pure concordant states are just the usual pure product state
and belong to a 4-manifold $\mathcal{P}_{\mathcal{C}}$. \ Expressed in terms
of density matrices, any state of this type is classical-classical with a
single product of projections operators, i.e. its density matrix is of the
form $\rho =|\Psi _{a}\rangle \langle \Psi _{a}|\otimes |\Psi _{b}\rangle
\langle \Psi _{b}|$.

In what follows, we will often refer to the set $\mathcal{C}$ rather than $%
\mathcal{C}_{-},$ since many of our considerations do not depend on the fact
that $\mathcal{C}$ is not itself a manifold structure; $\mathcal{C}$ and $%
\mathcal{C}_{-}$ differ by only a set of measure zero. 

\subsection{Parameterization of $\mathcal{C}$ (calculus on $\mathcal{C}$)}

We may calculate the 9 tangent vectors, namely $\{\overrightarrow{t}%
_{i}=\partial f/\partial x_{i}\}$; $i=1$ to $9$, where $x_{1}=\theta ,$ $%
x_{2}=\phi $ are the spherical polar coordinates for $\vec{m},$ $x_{3}=p,$ $%
x_{4}=n_{01},$ etc. \ The explicit forms of the $\overrightarrow{t}_{i}\in 
\mathbb{R}
^{15}$ are in appendix~\ref{a:15tangents}. \ We show there that these 9
tangent vectors form a linearly independent set almost everywhere in $%
\mathcal{C}$, i.e., that if there exists a set of real numbers $%
\{c_{1},c_{2},c_{3},...,c_{9}\}$ such that $c_{1}\overrightarrow{t}_{1}+c_{2}%
\overrightarrow{t}_{2}+c_{3}\overrightarrow{t}_{3}+...+c_{9}\overrightarrow{t%
}_{9}=\overrightarrow{0}$ then $c_{1}=c_{2}=...=c_{9}=0$. \ This procedure
fails when any of the $\vec{t}_{i}$ vanish. \ This occurs at the purely
coordinate singularities $\theta =0$ and $\theta =\pi $, which are not truly
singular points. \ It also happens at points with $p=1/2,\vec{n}_{0}=\vec{n}%
_{1}$ and at the points $p=0.$ \ As we have seen above, these correspond to
real singular points. \ However, since they occupy a set of measure zero,
the parametrization with $\theta ,\phi ,p,\vec{n}_{0},\vec{n}_{1}$ can be
used for integration with the 9-surface element 
\begin{equation*}
dS=\sqrt{|g|}\ d\theta \ d\varphi \ dp...\ dn_{11}\ dn_{12}\ dn_{13}
\end{equation*}%
where $g$ is the appropriate matrix tensor 
\begin{equation*}
g=\left( 
\begin{array}{c}
g_{ij}%
\end{array}%
\right) .
\end{equation*}%
$g$ consists of elements $g_{ij}=\overrightarrow{t}_{i}\cdot \overrightarrow{%
t}_{j}.$ Most of the off-diagonal elements of g's are zero (see the full
matrix form in appendix~\ref{a:15tangents}). We obtain $\sqrt{|g|}\
=16p^{3}(1-p)^{3}\sin {\theta }\ \{\ \sum_{i=1}^{3}{\
[pn_{0i}-(1-p)n_{1i}]^{2}}+(1-2p)^{2}\ \}$. %

\subsection{2-dimensional and 3-dimensional cross sections of $\mathcal{C}$}

It is difficult to visualize a 9-dimensional structure such as $\mathcal{C}$%
. \ Accordingly, we consider  sections of $\mathcal{C}$: intersections of $%
\mathcal{C}$ with coordinate planes obtained by setting some coordinates of $%
\mathbb{R}
^{15}$ equal to zero. \ In particular, we will consider 2-sections for which
13 coordinates are zero, and 3-sections for which 12 coordinates are zero. \
This will help to make clear the differences between entanglement and
discord. \ 
Because of the fact that $\mathcal{M}$ and $\mathcal{S}$ are convex
15-dimensional sets that include the origin, the 2-sections of $\mathcal{M}$
and $\mathcal{S}$ are all 2-dimensional convex sets. In fact all 2-sections
of $\mathcal{M}$ are either squares or disks centered at the origin\textit{\ 
}\cite{Zhou1}. \ (Note that using a different basis, such as the Gell-Mann
matrices \cite{Jakobczyk}, can result in the presence of other types of
geometry for the 2-sections such as triangles and parabolas.) \ Zhou \textit{%
et al.} \cite{Zhou1} were able to show that the occurrence of squares and
disks is determined by the commutativity properties of the operators
corresponding to the two axes: squares for commuting operators and disks for
non-commuting operators. \ Since the shape of $\mathcal{M}$ is determined by
positivity conditions on the eigenvalues; this is not so surprising: the
contribution of the coefficients $N_{ij}$ add in quadrature to the
eigenvalues of the non-commuting case. \ 

Making a complete survey of the 2-sections of $\mathcal{C}$ reveals
interesting similarities and differences to those of $\mathcal{M}$, as shown
in Tab.~(\ref{tab:2Ddiscord}). \ There are three \ geometries observed for $%
\mathcal{C}$: the square, the disk, and the cross. \ The first two are the
same as for $\mathcal{M},$ and presumably reflect similar physics, but the
cross is new and it occupies about one third of the table. \ It is the union
of the two line intervals $[-1,1]$ lying in the two Cartesian axes. \ This
is a locally 1-dimensional object (except at the origin, where the
intersection of the intervals occurs), which reflects the lower
dimensionality of $\mathcal{C}$, as compared to $\mathcal{S}$ or $\mathcal{M}%
.$ \ Furthermore, unlike entangled states, there are discordant states
arbitrarily close to the origin.

Using the explicit form for the 15 components of $N$'s for a concordant
state as expressed in Eqs.~(\ref{eq:mu1}) and~(\ref{eq:mu3}), \ the disk and
square of $\mathcal{C}$ are always specified with 2 independent variables
while this is not possible for the cross; the only two nonzero components of
the intersecting plane of the state cannot be nonzero at the same time if we
are to have zero discord.

An explicit example for the square is the state $\rho =\frac{I}{4}+\frac{1}{4%
}N_{10}\sigma _{1}\otimes \sigma _{0}+\frac{1}{4}N_{13}\sigma _{1}\otimes
\sigma _{3}$, $I$ is the identity matrix. This is a concordant subset of $%
\mathcal{M}$ obtained when $n_{02}=n_{03}=n_{12}=n_{13}=m_{2}=m_{3}=0$, $%
m_{1}=\pm 1$, and $n_{01}$ ($=n_{11}$) and $p$ are freely chosen from $%
\mathcal{A}$ such that $N_{10}=\pm (2p-1)$ and $N_{13}=\pm 2pn_{01}$. \ 

For the cross geometry, consider the example $\rho =\frac{I}{4}+\frac{1}{4}%
N_{10}\sigma _{1}\otimes \sigma _{0}+\frac{1}{4}N_{21}\sigma _{2}\otimes
\sigma _{1}$. The states of this set are discordant everywhere except on the
coordinate axes. \ The concordant states have only a single nonzero
component, either of $N_{10}=(2p-1)m_{1}$ and $N_{21}=2pm_{2}n_{01}$.
Specifically, in order for both $N_{10}$ and $N_{21}$ to be nonzero, the
product $2(2p-1)m_{1}m_{2}pn_{01}\neq 0$. \ \ But this implies that $%
N_{11}=2m_{1}pn_{01}\neq 0,\ N_{20}=(2p-1)m_{2}\neq 0$.

Let us consider the positions of the cross geometry in more detail, since
this geometry is unique to discord. \ States in a 2-section have the form%
\begin{equation*}
\rho =\frac{I}{4}+\frac{1}{4}N_{ij}\sigma _{i} \otimes \sigma _{j}+\frac{1}{4%
}N_{kl}\sigma _{k} \otimes \sigma _{l}
\end{equation*}%
so that we can refer to the $ij,kl$ section with $0\leq i,j,k,l\leq 3$. \ 
We first note that crosses occur only when at least one correlation function
is involved, i.e., at most one of the $j,l$ can be zero. [This observation
is related to the fact that we have considered the \textquotedblleft left"
discord measure, which is, in this case, on qubit A. \ Similar statements
hold for $i,k$ for the \textquotedblleft right" discord measure.] \ This is
expected, since discord requires correlation. \ An example of a density
matrix with cross behavior is 
\begin{equation*}
\rho =\frac{I}{4}+\frac{1}{4}N_{10}\sigma _{x} \otimes \sigma _{0}+\frac{1}{4%
}N_{21}\sigma _{y} \otimes \sigma _{x}.
\end{equation*}%
When $0<\left\vert N_{10}\right\vert <<1$ and $0<\left\vert
N_{21}\right\vert <<1$ , this state is separable but discordant. \ This
emphasizes the fact that discord, as compared to entanglement, is much more
resistant to dephasing, since states of this kind can be arbitrarily close
to the origin where the system is completely dephased. \ This state contains
quantum correlation because it combines the non-commuting operators $\sigma
_{x}\sigma _{0}$ and $\sigma _{y}\sigma _{x}.$ \ The choice of how to
measure qubit 1 (along the x-axis or along the y-axis) can have some effect
on how much information we gain about qubit 2. \ Finally, we look at the
case when all of the $i,j,k,l$ are nonzero. \ Crosses occur if and only if $%
i\neq k$ and $j\neq l,$ e.g., 
\begin{equation*}
\rho =\frac{I}{4}+\frac{1}{4}N_{11}\sigma _{x} \otimes \sigma _{x}+\frac{1}{4%
}N_{23}\sigma _{y} \otimes \sigma _{z}
\end{equation*}%
with $0<\left\vert N_{11}\right\vert <<1$ and $0<\left\vert
N_{23}\right\vert <<1$ is separable but discordant, but 
\begin{equation*}
\rho =\frac{I}{4}+\frac{1}{4}N_{11}\sigma _{x} \otimes \sigma _{x}+\frac{1}{4%
}N_{12}\sigma _{x} \otimes \sigma _{y}
\end{equation*}%
with $0<\left\vert N_{11}\right\vert <<1$ and $0<\left\vert
N_{12}\right\vert <<1$ (a disk state) is separable and concordant. \ It
seems that since measuring qubit 1 along the y or z axes gives no
non-trivial information, the choice involved does not generate discord.%

An examination of the 3-sections of $ \mathcal{C}$ is also revealing. Such
a state is of the form $\rho =\frac{I}{4}\ +\frac{1}{4}N_{ij}\sigma
_{i}\otimes \sigma _{j}\ +\frac{1}{4}N_{kl}\sigma _{k}\otimes \sigma _{l}\ +%
\frac{1}{4}N_{mn}\sigma _{m}\otimes \sigma _{n}$. \ Since three nonzero
coefficients are necessary to form a maximally entangled (pure) state, the
3-sections bring in qualitatively new physics.\ If the 3-section does
include maximally entangled states, then these states occupy the vertices of
a tetrahedron geometry, as shown previously for the Bell states\cite{Caves}. 
We show that the 3-sections can have zero or nonzero 3-volume. Using this
fact and the table of 2-section geometries [Tab.~(\ref{tab:2Ddiscord})] we
can characterize all allowed 3-section geometries. 
\begin{table}[tbp]
\begin{center}
\begin{tabular}{|cc||ccccccccccccccc|}
\hline
&  & 1 & 2 & 3 & 4 & 5 & 6 & 7 & 8 & 9 & 10 & 11 & 12 & 13 & 14 & 15 \\ 
\hline\hline
0X & 1 &  &  &  &  &  &  &  &  &  &  &  &  &  &  &  \\ 
0Y & 2 & D &  &  &  &  &  &  &  &  &  &  &  &  &  &  \\ 
0Z & 3 & D & D &  &  &  &  &  &  &  &  &  &  &  &  &  \\ 
X0 & 4 & S & S & S &  &  &  &  &  &  &  &  &  &  &  &  \\ 
Y0 & 5 & S & S & S & D &  &  &  &  &  &  &  &  &  &  &  \\ 
Z0 & 6 & S & S & S & D & D &  &  &  &  &  &  &  &  &  &  \\ 
XX & 7 & S & D & D & S & + & + &  &  &  &  &  &  &  &  &  \\ 
XY & 8 & D & S & D & S & + & + & D &  &  &  &  &  &  &  &  \\ 
XZ & 9 & D & D & S & S & + & + & D & D &  &  &  &  &  &  &  \\ 
YX & 10 & S & D & D & + & S & + & D & + & + &  &  &  &  &  &  \\ 
YY & 11 & D & S & D & + & S & + & + & D & + & D &  &  &  &  &  \\ 
YZ & 12 & D & D & S & + & S & + & + & + & D & D & D &  &  &  &  \\ 
ZX & 13 & S & D & D & + & + & S & D & + & + & D & + & + &  &  &  \\ 
ZY & 14 & D & S & D & + & + & S & + & D & + & + & D & + & D &  &  \\ 
ZZ & 15 & D & D & S & + & + & S & + & + & D & + & + & D & D & D &  \\ \hline
\end{tabular}%
\end{center}
\caption{Possible geometries of the 2-sections of the concordant subset $%
\mathcal{C}$. D and S stand for disk and square, respectively. Crosses
denote the union of two intervals on the two Cartesian axes.}
\label{tab:2Ddiscord}
\end{table}
\begin{figure*}[tph]
\caption{Two possible 3-sections of the concordant set $\mathcal{C}$ with
nonzero 3-volume. \ On the left is a union of 2 cones (case 3 below), while
on the right is a more complex, less easily characterized object (case 4
below) \ }
\label{fig3D_cone}\centering   
\begin{tabular}{cc}
\includegraphics[width=60mm]{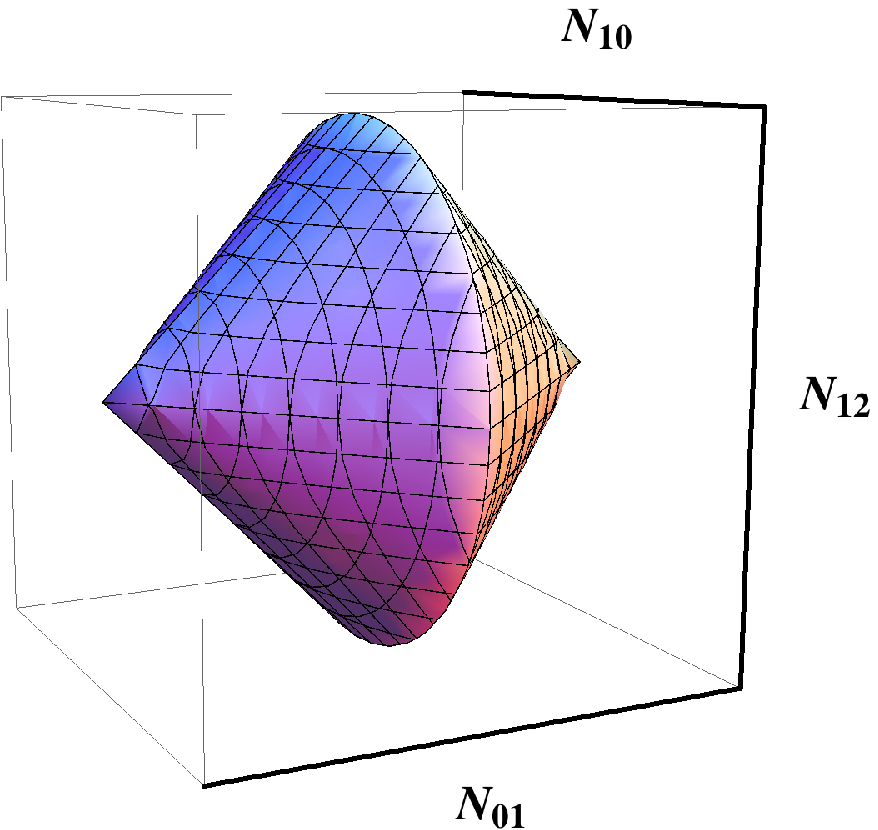} & %
\includegraphics[width=60mm]{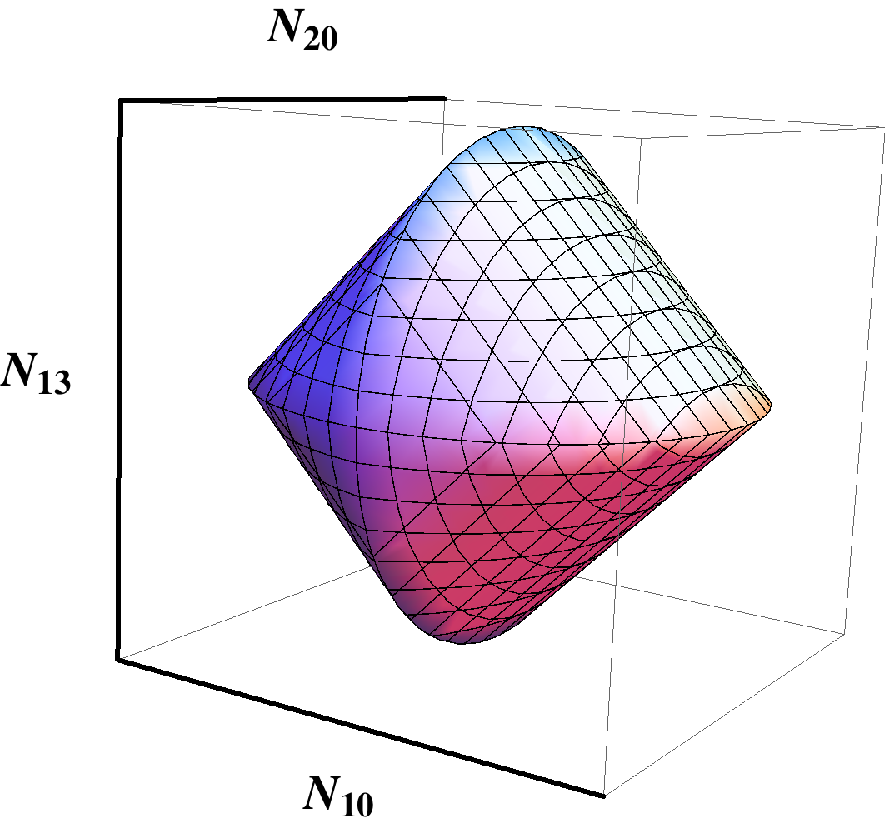}%
\end{tabular}%
\end{figure*}
%

When a 3-section has nonzero volume, it is specified by a set of three
independent parameters drawn from $\mathcal{A}$.\ As a relatively simple
example, consider the 3-section obtained by varying $~p, n_{01},$ and $%
n_{11},$ setting $m_{2}=m_{3}=n_{02}=n_{03}=n_{12}=n_{13}=0$ and $m_{1}=1$.
\ The 3 remaining parameters $p,n_{01}$, and $n_{11}$ are arbitrary.
Consider $m_{1}=1$. \ Then we find that the allowed values of $%
N_{01}=pn_{01}+(1-p)n_{11}$, $N_{10}=(2p-1)$, $N_{11}=pn_{01}-(1-p)n_{11}$
form a tetrahedron. \ Note that $N_{01}, N_{10}$, $N_{11}$ defined in this
way are independent of one another: the subclass of states having the
density matrix $\rho =\frac{1}{4}(\sigma _{0}\otimes \sigma
_{0}+N_{01}\sigma _{0}\otimes \sigma _{1}+N_{10}\sigma _{1}\otimes \sigma
_{0}+N_{11}\sigma _{1}\otimes \sigma _{1})$ contains all physical states in
a tetrahedron that is identical to the concordant tetrahedron specified
before by the set of parameters $\{p, n_{01}, n_{11}\}$.

We now make the observation that the 2-sections of this tetrahedral
3-section obtained by intersection with the $\left\{ N_{01},N_{10}\right\}
,\left\{ N_{01},N_{11}\right\} $ and $\left\{ N_{10},N_{11}\right\} $ planes
are all squares [see Tab.~(\ref{tab:2Ddiscord})]. \ This is in fact true of
any combination of \{$N_{0i},N_{j0},N_{ji}$\} (9 combinations in total) that
forms a tetrahedron of concordant states and leads to the classification of
all nonzero-volume 3-sections into 4 types.

1. \ Tetrahedron when all 3 2-sections are squares.

2. \ Unit ball when all 3 2-sections are disks. \ For example, the
combination $\{N_{01},N_{02},N_{03}\}$ by setting $p=1/2$, $\overrightarrow{n%
}_{0}=\overrightarrow{n}_{1}$ is in this class. Now, $N_{0i}=n_{0i}$
independent of $\overrightarrow{m}$. Similarly to the tetrahedron case
discussed above, all three components are independent of one another, which
indicates that any physical state made of three component $N_{0i}$ is a
concordant state. This property holds for other combinations such as $%
\{N_{10},N_{20},N_{30}\}$ by setting $\overrightarrow{n}_{0}=\overrightarrow{%
n}_{1}=0$ or $\{N_{11},N_{12},N_{13}\}$ by setting $p=1/2$, $\overrightarrow{%
n}_{0}=-\overrightarrow{n}_{1}$, and $m_{2}=m_{3}=0$, etc.

3. \ Union of 2 cones when 1 2-section is a square and 2 are disks.\ The
object can be thought of as 2 cones glue together at their bases or as the
surface of revolution formed when a square is rotated about an axis that
passes through its center.\ See Fig.~\ref{fig3D_cone} (left).

4. \ A less easily described 3-dimensional object shown in Fig.~\ref%
{fig3D_cone} (right), when the 1 2-section is a disk and 2 2-sections are
squares. \ For example, the combination $\{N_{01},N_{10},N_{12}\}$ obtained
by setting $m_{2}=m_{3}=n_{03}=n_{13}=0$ and $pn_{01}=(1-p)n_{11}$ and $%
pn_{02}=-(1-p)n_{12}$. \ This 3-section has 2-sections are a set of 2
squares [\{$N_{01},N_{10}$\} and \{$N_{01},N_{12}$\}] and 1 disk [\{$%
N_{10},N_{12}$\}], etc.

Last, we consider 3-sections with zero 3-volume. \ Its 2-sections include at
least one cross. \ Analytically, such a 3-section is specified by a union of
sets of equations and each set has at most 2 independent variables. \ An
example is the 3-section with nonzero \{$N_{01},N_{10},N_{21}$\} obtained by
setting either 1) $m_{2}=m_{3}=n_{02}=n_{03}=n_{12}=n_{13}=0$ and $%
pn_{01}=(1-p)n_{11}$ i.e. by at most 2 independent parameters ($p,n_{01}$)
or ($p,n_{11}$) or 2) $m_{1}=m_{3}=n_{02}=n_{03}=n_{12}=n_{13}=0$ and $p=1/2$
i.e. by 2 independent parameters $(n_{01},n_{11})$. \ \ If there are 3
crosses among the 2-sections, then the 3-section is locally 1-dimensional. \
The extreme example is the Bell-diagonal state with nonzero \{$%
N_{11},N_{22},N_{33}$\} that has a 3-section that is the union of 3
1-section objects - the coordinate axes.

\section{Time Evolution of Discord}

We now turn to the consequences of the topological analysis for the time
evolution of the quantum discord in 2-qubit systems. We are mainly
interested in decoherence, so we will assume that the initial state of the
system has finite discord that decreases overall, though perhaps not
monotonically, as time increases. \ The opposite behavior is obviously
possible: take a 2-qubit system in the fully mixed state and let it relax to
an entangled ground state by reason of contact with a cold bath. \ 

Hence we consider functions $D_{G}\left( \vec{N}\left( t\right) \right) ,$
where $\vec{N}$ is the 15-dimensional real generalized Bloch vector and $%
D_{G}\ $\ is the geometric discord. \ We will further assume that the system
tends to a limit as the time approaches infinity: $\lim_{t\rightarrow \infty
}\vec{N}\left( t\right) =\vec{N}_{\infty }$ and $D_{G\infty }=D_{G}\left( 
\vec{N}_{\infty }\right) .~$ If there are no self-intersections, the
trajectory $\left\{ \vec{N}\left( t\right) |0\leq t<\infty \right\} $ itself
is a 1-dimensional manifold. \ 

We briefly review the analysis of evolution entanglement. \ For any
evolution, we define the set of times when the entanglement vanishes: \ $%
T_{0}^{S}=\left\{ t|C\left( \vec{N}\left( t\right) \right) =0\right\} ,~$
where $C$ is the concurrence\cite{Bennett2,Wootters}. \ In previous work\cite%
{Zhou2}, transversality theory was applied to the intersections of
trajectories with the set $\mathcal{S}$ to analyze the possible forms of $%
T_{0}^{S}.$ \ It was found that when the trajectory and $S$ are transversal,
(the generic case), then entering behavior, $\mathsf{E}$, [entanglement
sudden death, $T_{0}^{S}=(t_{c},\infty $)] or oscillating, $\mathsf{O}$, ($%
T_{0}^{S}$ a union of finite intervals) are the first two possible behaviors
of the entanglement. \ It follows from transversality theorems that they are
both stable under small perturbations. \ The other two behaviors occur when
transversality is violated, which requires a symmetry in the dynamics or
other special conditions - then we can get half-life (approaching), $\mathsf{%
A}$, behavior ($T_{0}^{S}=\varnothing )$ or bouncing, $\mathsf{B}$,
behavior, when $T_{0}^{S}$ is a collection of isolated points. \ These two
latter behaviors are unstable to small perturbations. \ \ 

\ The condition for transversality theorems to hold is that the sum of the
dimensions of the intersecting manifolds be at least as great as the
dimension of the underlying space, which holds for trajectories (which have
dimension 1) and $\mathcal{S}$ since $\dim \mathcal{S}+1=16>15=\dim M.$ \
Since $\dim \mathcal{C}=9<14=\dim M-1,$ we cannot use the same reasoning for
discord evolution. \ Let us define $T_{0}^{D}=\left\{ t|D_{G}\left( \vec{N}%
\left( t\right) \right) =0\right\} .$ \ If we assume that $\vec{N}\left(
t\right) $ has a continuous first derivative and that $D_{G\infty }=0,$
there are two possibilities: ($T_{0}^{S}=\varnothing ,$ the null set)
(``half-life") or $T_{0}^{D}$ is a collection of isolated points (bouncing
behavior), and neither of these categories is stable with respect to small
perturbations. \ They happen only as a result of particular choices, when
the discord has a non-trivial relationship to the dynamics. \ This can \
happen naturally - for example, the origin $\vec{N}=0$ belongs to $C$ and $%
\vec{N}_{\infty }=0$ for a system in contact with a bath at high
temperature. 

\subsection{Unitary Evolution}

\textbf{Introduction}

Having classified the various possibilities for the evolution of
entanglement and discord, we now turn to the question of the realization of
these evolutions in explicit models. \ In this regard, it is useful to
distinguish between unitary evolution of the density matrix and non-unitary
evolutions. \ This distinction is of course crucial for the experimental
investigation of all types of coherence: unitary time evolution is by
definition coherent overall, all correlation measures should be unchanged by
local unitary evolution, but the behavior of different correlation measures
under nonlocal unitary time evolution can help to understand the
distinctions between different measures.

\textbf{Ising model}\newline
The Ising Hamiltonian: 
\begin{equation}
{H^{\text{I}}}=J\sigma _{3}\otimes \sigma _{3}
\end{equation}%
generates a two-qubit unitary operator of the system of the form: 
\begin{eqnarray}
U &=&\text{exp}(-iJt\,\sigma _{3}\otimes \sigma _{3})  \notag
\label{eq:unitary_Ising} \\
&=&\left( 
\begin{array}{cccc}
e^{-iJt} & 0 & 0 & 0 \\ 
0 & e^{iJt} & 0 & 0 \\ 
0 & 0 & e^{iJt} & 0 \\ 
0 & 0 & 0 & e^{-iJt}%
\end{array}%
\right)  \notag \\
&=&C\,\sigma _{0}\otimes \sigma _{0}-i\,S\,\sigma _{3}\otimes \sigma _{3}
\end{eqnarray}%
where we use the abbreviations $S=\sin {(Jt})$ and $C=\cos {(Jt)}$. \ This
would be an appropriate Hamiltonian for well-separated superconducting flux
qubits with the rings lying in the same plane when the applied field is zero.

Under the unitary transformation (\ref{eq:unitary_Ising}), any initial state
of general form (\ref{eq:general_rho}) evolves as: 
\begin{eqnarray}
\rho(t) &=& \frac{1}{4} ( \sigma_0 \otimes \sigma_0 + N_{0i}(t) \sigma_0
\otimes \sigma_i  \notag \\
&& + N_{i0} (t) \sigma_i \otimes \sigma_0 + N_{ij} (t) \sigma_i \otimes
\sigma_j ).
\end{eqnarray}

There are 7 constants of motion: $%
N_{03},N_{30},N_{33},N_{12},N_{21},N_{11},N_{22}$. \ With these constraints,
any system that is initialized in an \textbf{X}-state has constant quantum discord in
this model. \ This relatively simple model appears to be the most
non-trivial model that has trivial dynamics for the discord for a reasonable
wide class of states: $D$ is completely independent of time for the
8-dimensional space of \textbf{X}-states. 

\textbf{Heisenberg Model}

A Heisenberg model with the presence of all XYZ terms i.e. $%
H^{H}=J\sum_{i=X,Y,Z}{\sigma _{i}\otimes \sigma _{i}}$ is appropriate for
electron spin qubits with overlapping wavefunctions, which will then feel
the exchange interaction. \ \ The unitary transformation generated by $H^{H}$
is of course much richer. \ The class of states with constant discord is the
3-dimensional space of Bell-diagonal states.

The remaining 8 components evolve under the unitary transformation (\ref%
{eq:unitary_Ising}) as: 
\begin{eqnarray}
\hspace{2cm}N_{01}(t) &=&N_{01}C_{3}+N_{32}S_{3}  \notag
\label{eq:Separable01_DQC1} \\
\hspace{2cm}N_{32}(t) &=&N_{32}C_{3}-N_{01}S_{3}  \notag \\
\hspace{2cm}N_{02}(t) &=&N_{02}C_{3}-N_{31}S_{3}  \notag \\
\hspace{2cm}N_{31}(t) &=&N_{31}C_{3}+N_{02}S_{3}
\end{eqnarray}%
\begin{eqnarray}
\hspace{2cm}N_{10}(t) &=&N_{10}C_{3}+N_{23}S_{3}  \notag
\label{eq:Separable02_DQC1} \\
\hspace{2cm}N_{23}(t) &=&N_{23}C_{3}-N_{10}S_{3}  \notag \\
\hspace{2cm}N_{20}(t) &=&N_{20}C_{3}-N_{13}S_{3}  \notag \\
\hspace{2cm}N_{13}(t) &=&N_{13}C_{3}+N_{20}S_{3}
\end{eqnarray}%
where {$S_{3}=\sin (2Jt)$} and {\ $C_{3}=\cos (2Jt)$}.

Note that under a unitary transition the purity of the state is conserved
i.e. $|\overrightarrow{N}(t)|^{2}=|\overrightarrow{N}|^{2}$. The two
separate groups with time dependent components of $\rho (t)$ in Eqs.~(\ref%
{eq:Separable01_DQC1}) and ~(\ref{eq:Separable02_DQC1}) are two groups of
DQC1 separable states.

For purposes of illustration we choose the initial condition such that only $%
N_{20}$ is nonzero and $N_{20}=1$ (a concordant state). \ \ The system
evolves as: 
\begin{eqnarray}
\hspace{2cm}N_{20}(t) &=&C_{3}  \notag  \label{eq:DQC1_unitary} \\
\hspace{2cm}N_{13}(t) &=&S_{3}
\end{eqnarray}%
and is a separable state i.e. $C(t)=0$ with concordant subset as the union
of $N_{20}(t)$ and $N_{13}(t)$ axes (see Tab.~\ref{tab:2Ddiscord}). Quantum
trajectory of the system is the unit circle ${N_{20}(t)}^{2}+{N_{13}(t)}%
^{2}=1$.

\begin{figure}[h]
\begin{center}
\vspace*{0.1cm} \includegraphics[width=7.5cm]{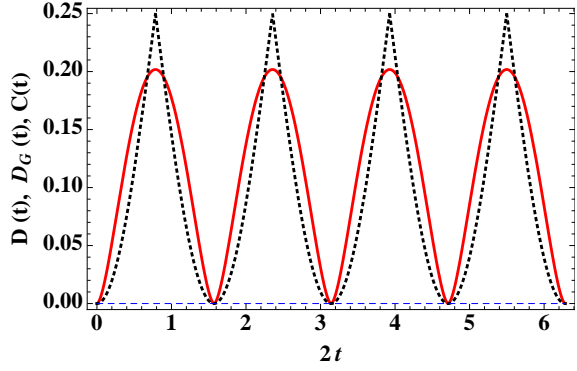}
\end{center}
\par
\vspace{-0.4cm}
\caption{(Color online) Time dependence of the quantum discord (red solid
line), geometric discord (black dotted line), and concurrence (blue dashed
line) of system described by (\protect\ref{eq:DQC1_unitary}). }
\label{fig:UdiscordZZ}
\end{figure}

The quantum discord of ~(\ref{eq:DQC1_unitary}) is (see detailed
calculations in Appendix \ref{ap:discord_calculation}): 
\begin{eqnarray}
D(t) &=&-\frac{1}{2}[(1+C_{3})\log (1+C_{3})+(1-C_{3})\log (1-C_{3})]  \notag
\\
&+&1-\frac{1}{2}[(1-S_{3})\log (1-S_{3})+(1+S_{3})\log (1+S_{3})]  \notag
\end{eqnarray}%
and the geometric quantum discord is: 
\begin{equation*}
D_{G}(t)=\frac{1}{4}(1-\max {\{C_{3}}^{2},{S_{3}}^{2}\})
\end{equation*}
which are shown in Fig.~\ref{fig:UdiscordZZ}. \ The entanglement is
identically zero, while the discord oscillates with maximum ($\approx 0.2$)
at $t=\frac{\pi }{16}+n\frac{\pi }{8}$ and vanishes at $t=n\frac{\pi }{8}$.
\ \ In the semiclassical picture, the two spins precess about one another. \
We have chosen a starting state that is separable, and the mutual precession
does not generate entanglement. \ This is true for nearly all separable
initial conditions, so our choice of initial state is fairly generic. \ For
the discord, however, the situation is quite different. 
\ To have zero
discord, the classical states of one subsystem need to pair up with the
mixed states of the other. \ This requires additional phase relations. \
Because these phase relations are oscillating, we get a periodic behavior of
the discord. \ Note that the geometric and quantum discord behave very
similarly, as is nearly always the case. \ The only significant distinction
is the linear (quadratic) zeros for the quantum (geometric) discord
corresponding to the linear and quadratic distance measures in the
definitions.

\textbf{Anisotropic XY-model} \newline
Consider an anisotropic exchange Hamiltonian with cross-product terms: 
\begin{equation}
{H^{XY}}=J_{xy}\sigma _{x}\otimes \sigma _{y}+J_{yx}\sigma _{y}\otimes
\sigma _{x}.
\end{equation}%
%
%
%
%
%
%
%
%
%
%
The corresponding unitary operator is: 
\begin{widetext}
\begin{eqnarray} \label{eq:unitary_XY}
U(t) &=& \text{e}^{-it(J_{xy} \, \sigma_x \otimes \sigma_y + J_{yx} \, \sigma_y \otimes \sigma_x) }    
\nonumber\\
      &=& \left( \begin{array}{cccc}
\cos{(J_{xy}+J_{yx})t } & 0 & 0                                                 & -\sin{ (J_{xy}+J_{yx}) t }          \\
0         & \cos{(J_{xy}-J_{yx}) t }     &\sin{ (J_{xy}-J_{yx}) t }     &0 \\
0         & -\sin{ (J_{xy}-J_{yx}) t }     & \cos{ (J_{xy}-J_{yx}) t }  &0  \\
\sin{ (J_{xy}+J_{yx}) t }         & 0 &   0                                        &\cos{ (J_{xy}+J_{yx}) t }  
\end{array} \right) \nonumber\\
& = & C_1\, C_2 \,\,\sigma_0 \otimes \sigma_0 - S_1\, S_2 \,\, \sigma_3 \otimes \sigma_3 -   i  \, ( S_1\,C_2  \,\, \sigma_1 \otimes \sigma_2 + S_2\,C_1 \,\,\sigma_2 \otimes \sigma_1 )
\end{eqnarray}
\end{widetext}
where $S_{1}=\sin {\ (J_{xy}t})$, $C_{1}=\cos {(J_{xy}t)}$ and $S_{2}=\sin {%
\ (J_{yx}t})$, $C_{2}=\cos {(J_{yx}t)}$. These cross-product terms reflect
the fact that the number of constants of motion decreases as compared to
that of the Ising model and we expect to see different evolution behaviors
for the quantum correlations in the Bell-diagonal class. Let us consider the
situation where $J_{xy}=-J_{yx}$. The unitary operator simplifies 
\begin{eqnarray}
U(t) &=&{C_{2}}^{2}\,\,\sigma _{0}\otimes \sigma _{0}+{S_{2}}^{2}\,\,\sigma
_{3}\otimes \sigma _{3}  \notag  \label{eq:unitary_X=-Y} \\
&&+i\,S_{2}\,C_{2}\,\,(\sigma _{1}\otimes \sigma _{2}-\,\,\sigma _{2}\otimes
\sigma _{1}).
\end{eqnarray}%
This model can arise from the Dzyaloshinskii-Moriya interaction between two
electron spins whose separation vector is along the z-axis. \ 
Consider the initial state of the general form in Eq. (\ref{eq:general_rho}%
). The state at time $t$ is given by: 
\begin{widetext}
\begin{eqnarray}
\hspace{0cm}N_{03}(t) &=& \frac{1}{2}    [  N_{03} + N_{30}  + ( N_{03}  - N_{30}) \cos{4 J_{yx} t}  + (N_{11} + N_{22}) \sin{4 J_{yx} t}              ]         \nonumber\\ 
\hspace{0cm}N_{30}(t) &=& \frac{1}{2}    [  N_{03} + N_{30}  + (-N_{03} + N_{30}) \cos{4 J_{yx} t}  - (N_{11} + N_{22}) \sin{4 J_{yx} t}              ]         \nonumber \\
\hspace{0cm}N_{11}(t) &=& \frac{1}{2}    [  N_{11} - N_{22}  + (-N_{03} + N_{30}) \sin{4 J_{yx} t}  + (N_{11} + N_{22}) \cos{4 J_{yx} t}              ]         \nonumber \\
\hspace{0cm}N_{22}(t) &=& \frac{1}{2}    [  -N_{11} + N_{22}  + (-N_{03} + N_{30}) \sin{4 J_{yx} t}  + (N_{11} + N_{22}) \cos{4 J_{yx} t} ]     
\end{eqnarray}
\end{widetext}
and 
\begin{eqnarray}
\hspace{-4cm}N_{33}(t) &=&N_{33}  \notag \\
\hspace{-4cm}N_{12}(t) &=&N_{12}  \notag \\
\hspace{-4cm}N_{21}(t) &=&N_{21}
\end{eqnarray}%
%
%
%
%
and 
\begin{eqnarray}
\hspace{-1cm}N_{01}(t) &=&N_{01}\cos {2J_{yx}t}-N_{13}\sin {2J_{yx}t}  \notag
\\
\hspace{-1cm}N_{13}(t) &=&N_{01}\sin {2J_{yx}t}+N_{13}\cos {2J_{yx}t} \\
\hspace{-1cm}N_{02}(t) &=&N_{02}\cos {2J_{yx}t}-N_{23}\sin {2J_{yx}t}  \notag
\\
\hspace{-1cm}N_{23}(t) &=&N_{02}\sin {2J_{yx}t}+N_{23}\cos {2J_{yx}t} \\
\hspace{-1cm}N_{10}(t) &=&N_{10}\cos {2J_{yx}t}+N_{31}\sin {2J_{yx}t}  \notag
\\
\hspace{-1cm}N_{31}(t) &=&-N_{10}\sin {2J_{yx}t}+N_{31}\cos {2J_{yx}t} \\
\hspace{-1cm}N_{20}(t) &=&N_{20}\cos {2J_{yx}t}+N_{32}\sin {2J_{yx}t}  \notag
\\
\hspace{-1cm}N_{32}(t) &=&-N_{20}\sin {2J_{yx}t}+N_{32}\cos {2J_{yx}t}.
\end{eqnarray}%
%
%
%
%
In this XY-model, the discord of the Bell-diagonal class of states is no
longer independent of time. In the \textbf{X}-type of class of \ states,
only the states with only three nonzero components $%
\{N_{12}(t),N_{21}(t),N_{33}(t)\}$ have time-independent discord. All
physical states of this type lie in the tetrahedron similar to the geometry
of the Bell-diagonal states with the concordant subset as the union of the
three intervals in the Cartesian axes.

A Werner state\cite{Werner}: 
\begin{eqnarray}
\rho (0)=\rho _{W}(0) &\equiv &\frac{1}{4}[\sigma _{0}\otimes \sigma
_{0}-\alpha \sum_{i}\sigma _{i}\otimes \sigma _{i}]  \notag
\label{eq:Werner_initial} \\
&=&\frac{1-\alpha }{4}I+\alpha |\Psi ^{-}\rangle \langle \Psi ^{-}|
\end{eqnarray}%
where $|\Psi ^{-}\rangle =\frac{|01\rangle -|10\rangle }{\sqrt{2}}$ and $%
0\leq \alpha \leq 1$ can exhibit sudden death/birth and oscillating behavior
unitarily in this model. It evolves as: 
\begin{eqnarray}
\begin{array}{lllll}
N_{03}(t) & =- & N_{30}(t) & = & -\alpha \sin {4J_{yx}t} \\ 
N_{11}(t) & = & N_{22}(t) & = & -\alpha \cos {4J_{yx}t} \\ 
N_{33}(t) & = &  &  & -\alpha .%
\end{array}
\label{eq:W_evolve}
\end{eqnarray}%
With the introduction of the mixing parameter $\alpha $ we can also find
transitions between different evolution categories. \ $\alpha =1$ is a pure
maximally entangled state, while $\alpha =0$ is the completely mixed state.
\ This Werner state is separable when $\alpha \leq \frac{1}{3}$ (Eq.~\ref{eq:Werner_initial}).\ As $\alpha $ is varied, we find in the Bloch vector
representation that $(N_{11},N_{22},N_{33})=(-\alpha ,-\alpha ,-\alpha ),$
so the vector lies on a line segment whose end points are the origin $\alpha
=0$ and a point on the boundary of $\mathcal{M}$.

\begin{figure*}[tbph]
\begin{center}
\vspace{0.1cm} \hspace{-0.2cm} \includegraphics[width=12.2cm]{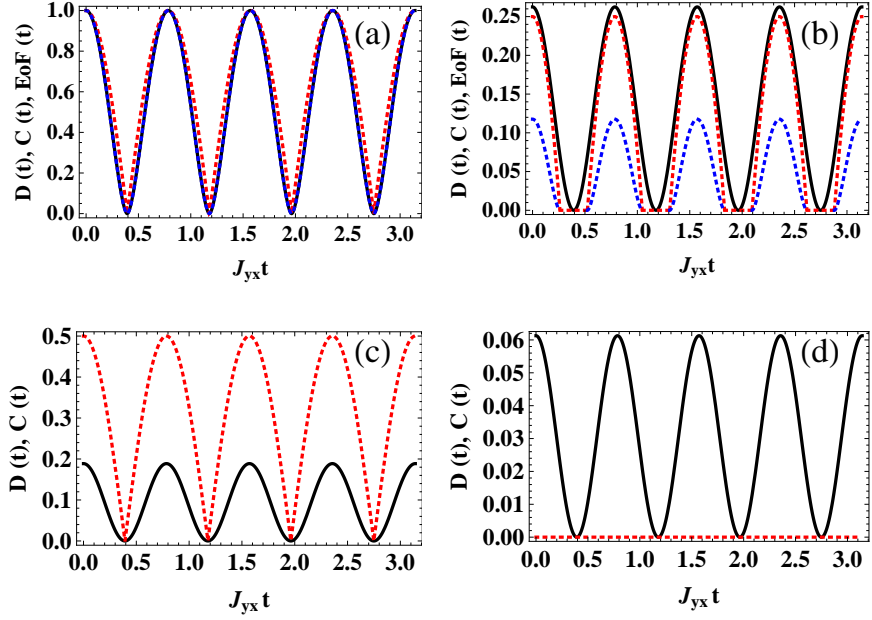}
\end{center}
\par
\vspace{-0.7cm}
\caption{(Color online) Time dependence of the quantum discord (black solid
lines), concurrence (red dotted lines), and entanglement (blue dashed
lines). \ For (a) and (b) the inital state is a Werner state with initial
conditions $\protect\alpha =1$ (a) and $\protect\alpha =1/2$ (b), as defined
in Eq. ~(\protect\ref{eq:Werner_initial}). \ The time evolution is given by
Eq. ({\protect\ref{eq:W_evolve}}). \ For (c) and (d) the initial state is a
Bell-diagonal state that evolves according to Eq. ({\protect\ref%
{eq:Bell_XY_model_xxyy2zz}}) with entangelement parameter $\protect\beta =1/2
$ (c) and $\protect\beta =1/4$ (d). }
\label{fig:pure_W}
\end{figure*}

If the initial condition is $\alpha =1$, the system evolves away from a
maximally entangled situation. \ The concurrence and quantum discord are 
$C(t)=|\cos {4J_{yx}t}|$ 
and 
\begin{widetext}
\begin{eqnarray}
\mathcal{D} (t) &=& 1 - \frac{1}{2}  \{    (\cos{2 J_{yx} t} + \sin {2 J_{yx} t})^2 \log[  (\cos{2 J_{yx} t} + \sin {2 J_{yx} t})^2   ] \nonumber\\
&& +  (\cos{2 J_{yx} t}  - \sin {2 J_{yx} t})^2 \log[  (\cos{2 J_{yx} t} - \sin {2 J_{yx} t})^2   ]	        \}.
\end{eqnarray}
\end{widetext}
The peaks of $D\left(t\right)$ correspond to the two pure-state points of
the 2 maximally entangled Bell states: $N_{11}[t=(2n+1)\frac{\pi }{4}%
]=N_{22}(t)=-N_{33}=1$ and $N_{11} [t=(2n)\frac{\pi }{4}] = N_{22} (t) =
N_{33} = -1$ while the vanishing discord points are two of the 4 pure-state
points of the concordant tetrahedron $\{N_{30}, N_{03}, N_{33}\}$ (see Fig.~%
\ref{fig:pure_W}). Since the entanglement and the discord vanish at discrete
points, this is $\mathsf{\ BB} $ joint evolution of entanglement and
discord.
%
When the initial condition is $\alpha =1/2,$ then the state starts as
partially entangled. \ Evolution under the XY-Hamiltonian leads to
entanglement death [$C=0$ in the region $\frac{\pi }{12}+n\frac{\pi }{4}\leq
t\leq \frac{\pi }{6}+n\frac{\pi }{4}$] and rebirth as illustrated in Fig.~%
\ref{fig:pure_W}

$C$ can be computed explicitly as 
$C(t)=\max {\ \{0,\Delta _{\lambda }\}}$ 
with 
$\Delta _{\lambda }=\frac{1}{8}[\sqrt{9+4\cos {8J_{yx}t}+2\Delta }-\sqrt{%
9+4\cos {8J_{yx}t}-2\Delta }-2]$ 
and $\Delta =\sqrt{2\cos {16J_{yx}t}+18\cos {8J_{yx}t}+16}$. 
This is $\mathsf{O}$-type behavior.

The quantum discord in this case is: 
\begin{eqnarray}
D(t) &=&\frac{5}{8}\log {5}-\frac{3-2\sin {4J_{yx}t}}{8}\log {(3-2\sin {%
4J_{yx}t})}  \notag \\
&&-\frac{3+2\sin {4J_{yx}t}}{8}\log {(3+2\sin {4J_{yx}t})}.
\end{eqnarray}%
$D\left( t\right) $ vanishes at the discrete points $t=\frac{\pi }{8}+n\frac{%
\pi }{4}$ as shown in Fig.~\ref{fig:pure_W}(b). This is $\mathsf{B}$-type
behavior. \ These concordant points belong to the interior of the above
tetrahedron. 
Thus the joint entanglement-discord evolution is of type $\mathsf{OB}.$

This Werner class of states belongs to a more general class: the
Bell-diagonal type: 
\begin{equation*}
\rho (0)=\rho _{\text{Bell}}(0)\equiv \frac{1}{4}[\sigma _{0}\otimes \sigma
_{0}+\sum_{i}(N_{ii}\sigma _{i}\otimes \sigma _{i})].
\end{equation*}%
with arbitrary values for $N_{11},N_{22},$ and $N_{33}$, each within the
range $[-1,1]$. \ In this larger class we find other types of joint
evolution. \ For example, consider the initial state as the Bell-diagonal
subclass with constraint $N_{11}=N_{22}=-\frac{N_{33}}{2}=\beta $ with $%
0\leq \beta \leq 1/2$. The time dependent Bloch vector becomes: 
\begin{eqnarray}
\begin{array}{lllll}
N_{03} (t) & = & -N_{30}(t) & = & \beta \sin{4 J_{yx} t } \\ 
N_{11}(t) & = & N_{22}(t) & = & \beta \cos{4 J_{yx} t} \\ 
N_{33}(t) & = & N_{33}. &  &  \\ 
&  &  &  & 
\end{array}
\label{eq:Bell_XY_model_xxyy2zz}
\end{eqnarray}
The initial concurrence of this Bell-diagonal state is $C(0)=\max
\{0,(4\beta -1)/2\}$ which implies that it is partially entangled for $%
1/4<\beta \leq 1/2$. The concurrence evolves as: 
$C(t)=\max \{0,\Delta _{B}\}$ 
where 
\begin{eqnarray}
\Delta _{B} &=& \frac{1}{4} [\sqrt{1+4\beta +8\beta ^{2}{\cos {4J_{yx}t}}%
^{2} + 4\beta {\ \Gamma _{B}}}  \notag \\
&& -\sqrt{1+4\beta +8\beta ^{2}{\cos {4J_{yx}t}}^{2}-4\beta {\ \Gamma _{B}}}
\notag \\
&&-2(1-2\beta )]  \notag
\end{eqnarray}%
with $\Gamma _{B}=\sqrt{{\cos {(4J_{yx}t)}}^{2}(1+4\beta +4\beta ^{2}{\cos {%
(4J_{yx}t)}}^{2})}$; while the quantum discord in this case is given by: 
\begin{eqnarray}
\mathcal{D}(t) &=&\frac{5}{8}\log {5}-\frac{3-2\sin {4J_{yx}t}}{8}\log {%
(3-2\sin {4J_{yx}t})}  \notag \\
&&-\frac{3+2\sin {4J_{yx}t}}{8}\log {(3+2\sin {4J_{yx}t})}.
\end{eqnarray}%
$D\left( t\right) $ vanishes at the discrete points $t=\frac{\pi }{8}+n\frac{%
\pi }{4}$ as shown in Fig.~\ref{fig:pure_W}(b). \ For $\beta =1/2$ the
concurrence reduces to a simpler form $C(t)=\frac{|{\cos {4J_{yx}t}|}}{2}$\
and the quantum discord and quantum entanglement evolve in a relatively
similar manner (bouncing) so that the joint evolution is again $BB$ as seen
in Fig.~\ref{fig:pure_W}(c). \ The Bloch vector has the time dependent form
given in Eq.~(\ref{eq:Bell_XY_model_xxyy2zz}). When $\beta =1/4$ we again
have zero entanglement coexisting with oscillating discord as seen in Fig.~%
\ref{fig:pure_W}(d)\ 

\subsection{\protect\bigskip Non-unitary evolution}

\textbf{Ising model with random telegraph noise}

Now we study how the system decoheres using a minimal random telegraph noise
model: an unbiased single-fluctuator random telegraph noise, in a Markovian
or/and non-Markovian process subject to an applied magnetic field along the $%
z$-direction. Such a system can be described by the following Hamiltonian: 
\begin{eqnarray}  \label{eq:Ising_RTN}
H=H^{I} + H^{RTN} + H^{Z}
\end{eqnarray}
where ${H^{I}}= J\, \sigma_z \otimes \sigma_z, $ $H^{RTN}=s(t)\, g_z\,
\sigma_z \otimes \sigma_0 $, and the Zeeman energy 
$H^{Z}=B_z \,\sigma_0 \otimes \sigma_z$. 

Using the quasi-Hamiltonian method (cf.~Refs.\cite{Cheng,Bob}) this problem
can be solved exactly.

\begin{equation*}
\overrightarrow{N}(t)=\langle f|e^{-iH_{q}t}|i\rangle \overrightarrow{N}_{0}
\end{equation*}%
where $|f\rangle \equiv |i\rangle =(1,1)^{T}/\sqrt{2}$ 
and 
\begin{equation*}
H_{q}=i\lim_{\Delta {t}\rightarrow 0}{\ \frac{\mathbf{{\Gamma }-\mathbf{I}%
_{30}}}{\Delta {t}}}
\end{equation*}%
where 
\begin{equation*}
\mathbf{{\Gamma }=\left( 
\begin{array}{cc}
(1-\gamma \Delta {t})\mathbf{T}_{0} & \gamma \Delta {t}\mathbf{T}_{0} \\ 
\gamma \Delta {t}\mathbf{T}_{1} & (1-\gamma \Delta {t})\mathbf{T}_{1}%
\end{array}%
\right) }
\end{equation*}%
and $\mathbf{T}_{0}=e^{-i\Delta {t}H[s(t)=1]}$ and $\mathbf{T}%
_{1}=e^{-i\Delta {t}H[s(t)=-1]}$ correspond to the temporal transfer matrix
when the noise sequence switches from $s=1$ to $s=-1$, respectively, with
transition rate $\gamma $. $H_{q}$ is called the time-independent
quasi-Hamiltonian\cite{Bob}. Eigenvalues of $H_{q}$ can contain imaginary
numbers which give decay rates. Real parts of its eigenvalues appear in the
oscillation frequencies. 
\begin{figure*}[tbph]
\begin{center}
\vspace{0.1cm} \hspace{-0.2cm} \includegraphics[width=17.2cm]{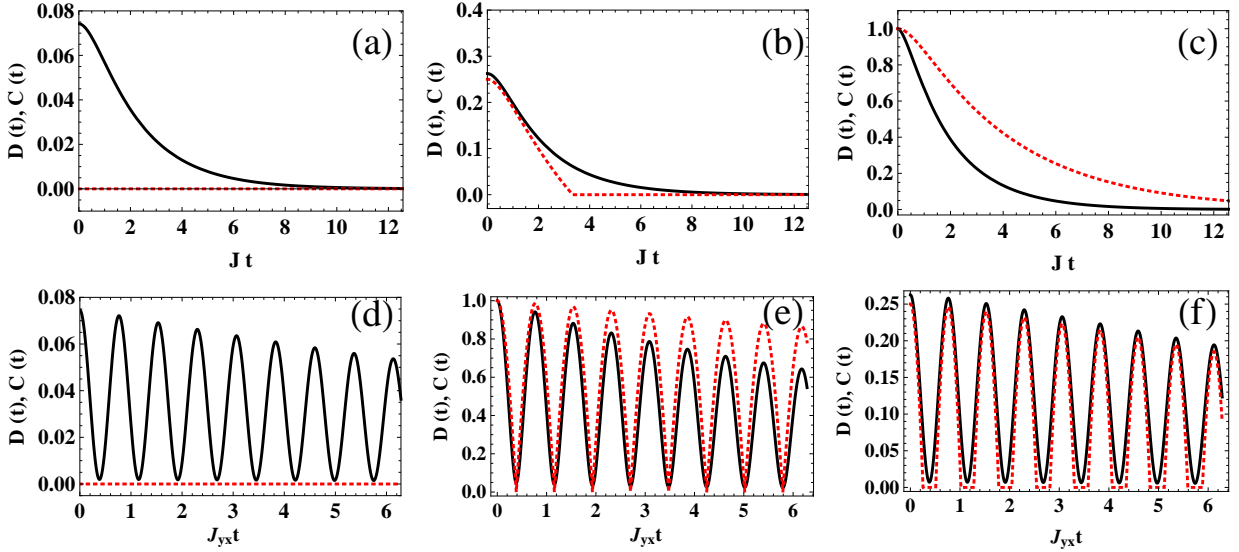}
\end{center}
\par
\vspace{-5.7cm}
\caption{(Color online) Six possible categories for the joint evolution of
quantum discord (solid lines) and entanglement (dashed lines).\ The initial
state is the Werner state defined in Eq.~(\protect\ref{eq:Werner_initial})
[with $\protect\alpha =1/4$ for (a) and (d), $\protect\alpha =1/2$ for (b)
and (e), and $\protect\alpha =1$ for (c) and (f)] . \ The Markovian
evolution is characterized by the paramters $\protect\gamma %
=J=J_{yx}=3g_{z}=3B_{z}=1$. Top panels are results obtained using the Ising
Hamiltonian and bottom panels are from the XY-model. }
\label{fig:Mar_cats}
\end{figure*}
Note that all the three terms in (\ref{eq:Ising_RTN}) are mutually commuting
so that the solution for the entire system can be obtained by solving each
single-term Hamiltonian separately (see Appendix \ref{ap:uncorrelated_noise}%
). Above we obtained full closed forms for all Bloch vector components of
the Ising model as the system evolves unitarily where the \textbf{X}-type
class has all 7 components as constants of motion. This means that all these
components are affected only by the noise and applied field parts of $H$. As
a consequence, all states of \textbf{X}-type exhibit only categories $%
\mathsf{A}$ (for discord) and $\mathsf{A}$ and $\mathsf{E}$ (for
entanglement) in Markovian regime. \ If the initial state is outside the 
\textbf{X}-type [union of the two subclasses (\ref{eq:Separable01_DQC1}) and
(\ref{eq:Separable02_DQC1})], its evolution type depends on $J$. \ For
example, Eq.~(\ref{eq:DQC1_unitary}) now becomes 
\begin{eqnarray}  \label{eq:Bloch_RTN_ZZ}
N_{20}(t) &=&e^{-\gamma \,t}\cos {(2Jt)}\,F(R_{0})  \notag \\
N_{13}(t) &=&e^{-\gamma \,t}\sin {(2Jt)}\,F(R_{0})
\end{eqnarray}
%
%
%
%
%
%
%
%
\begin{figure*}[tbph]
\begin{center}
\vspace{0.1cm} \hspace{-0.2cm} \includegraphics[width=17.2cm]{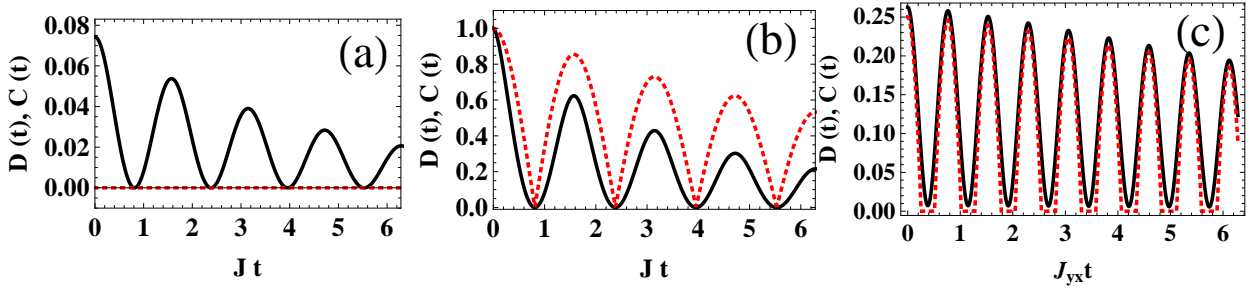}
\end{center}
\par
\vspace{-8.9cm}
\caption{(Color online) Three categories (seen in both Ising and XY-models)
for the joint evolution of quantum discord (solid lines) and entanglement
(dashed lines). \ The initial state is the Werner state defined in Eq.~(%
\protect\ref{eq:Werner_initial}) with $\protect\alpha =1/4$ for (a), $%
\protect\alpha =1$ for (b), and $\protect\alpha =1/2$ for (c). \ The
evolution is non-Markovian and is characterized by the parameters $\protect%
\gamma =0.27, J=J_{yx} = 3 g_{z} = 3 B_{z} = 1$. (a) and (b) are obtained
using the Ising model and (c) the XY-model.}
\label{fig:nonM_cats}
\end{figure*}
where $R_{0}=\sqrt{{g_{z}}^{2}-\gamma ^{2}/4}$ and 
\begin{equation}
F(R_{0})=\frac{2\,R_{0}\,\cosh {\ (2iR_{0}t)}-i\,\gamma \,\sinh {(2iR_{0}t)}%
}{(2\,R_{0})}.
\end{equation}%
Note that these components are independent of the applied field $B$. The
dynamical process is Markovian if $\gamma /2>g_{z}$ and non-Markovian if $%
\gamma /2<g_{z}$. The quantum discord [analytic form obtained in Appendix %
\ref{ap:DQC1_RTN}] of this system exhibits only category $\mathsf{B}$ and
entanglement is 0 for all $t$.

The time evolution of the Bloch vector of the Bell-diagonal state in this
model is: 
\begin{eqnarray}
N_{11}(t) &=&\,\,\,N_{11}\,{\ e^{-\gamma t}}\,\,F(R_{0})\,\cos {\ (2B_{z}t)}
\notag  \label{eq:NtBell_ZZ_RTN} \\
N_{12}(t) &=&-N_{11}\,{\ e^{-\gamma t}}\,\,F(R_{0})\,\sin {\ (2B_{z}t)} 
\notag \\
N_{21}(t) &=&N_{22}\,{\ e^{-\gamma t}}\,\,F(R_{0})\,\sin {\ (2B_{z}t)} 
\notag \\
N_{22}(t) &=&N_{22}\,{\ e^{-\gamma t}}\,\,F(R_{0})\,\cos {\ (2B_{z}t)} 
\notag \\
N_{33}(t) &=&N_{33}.
\end{eqnarray}%
%
%
%
%
%
%
%
%
This model yields a wide range of possible joint evolutions depending on
initial conditions. \ For the Werner state with $\alpha =1/4,$ we get zero
entanglement at all times, and the discord shows $\mathsf{A}$ behavior, as
shown in\ Fig.~\ref{fig:Mar_cats}(a). \ For $\alpha =1/2,$ we find $\mathsf{E%
}$-type (sudden death) behavior, while the discord shows $\mathsf{A}$
behavior. \ For $\alpha =1,$ we find $\mathsf{A}$-type behavior for both
types of correlation. \ These evolutions are all in the Markovian regime. \
We illustrate two categories $\mathsf{A}$ and $\mathsf{E}$ for quantum
correlations in Figs.~\ref{fig:Mar_cats}(a) - (c) for the Werner state (\ref%
{eq:Werner_initial}) in a Markovian regime. Discord at time $t$ of this
state is obtained in Appendix \ref{ap:discord_Werner_Ising_RTN}. In case $%
\alpha =1$: 
\begin{equation*}
\mathcal{D}(t)=1+\lambda _{+}\log {\ \lambda _{+}}+\lambda _{-}\log {\
\lambda _{-}}
\end{equation*}%
where $\lambda _{\pm }=\frac{1\pm \sqrt{{\ N_{11}(t)}^{2}+{\ N_{12}(t)}^{2}}%
}{2}$ and the corresponding concurrence is: 
\begin{eqnarray}
2C(t) &=&|{N_{11}}(t)|+\sqrt{1-{\ N_{12}(t)}^{2}}-\big{|}|{N_{11}}(t)| 
\notag \\
&&-\sqrt{1-{\ N_{12}(t)}^{2}}\big{|}.
\end{eqnarray}%
\textbf{\ XY-model with random telegraph noise } 
\begin{equation*}
H=H^{XY}+H^{RTN}+H^{Z}
\end{equation*}%
where ${H^{XY}}=J_{xy}\left( \sigma _{x}\otimes \sigma _{y}-\sigma
_{y}\otimes \sigma _{x}\right) .$ \ This is more complicated than the Ising
case (see Appendix \ref{ap:XY_Hq}) as the XY-term does not commute with the
noise and the B-field terms. Consider the initial state as Werner state (\ref%
{eq:Werner_initial}). The Bloch vector of the system evolves as: 
\begin{eqnarray}
N_{03}(t) &=&-N_{30}(t)  \notag \\
N_{12}(t) &=&- N_{21}(t)  \notag \\
N_{11}(t) &=&N_{22}(t)  \notag \\
N_{33}(t) &=&N_{33} \ (=-\alpha).
\end{eqnarray}%
$N_{12}(t)$ and $N_{21}(t)$ are addition elements when noise is added as
compared to the case without noise (see the corresponding unitary
transformation).

In the evolutions generated by the XY-Hamiltonian, oscillations occur in
the two correlation measures, and we find zero entanglement and $\mathsf{B}$
behavior for the discord for $\alpha =1/4,$ while $\alpha =1/2$ leads to $%
\mathsf{BB}$ behavior for the joint evolution, and $\alpha =1$ gives $%
\mathsf{OB}$ joint evolution, with the behavior of the entanglement given
first. \ The actual evolutions are shown in Figs.~\ref{fig:Mar_cats}(d),
(e), (f).

Note that the quantum discord never quite vanishes for this case of the
applied field on the second qubit $B_{z}$ which guarantee two components $%
N_{12}(t) = -N_{21}(t)\neq 0$ (which are zero in the case of the unitary
transformation). Recall that for the unitary evolution, the quantum discord
vanishes at $\frac{\pi }{12}+n\frac{\pi }{4}\leq J_{yx}t\leq \frac{\pi }{6}+n%
\frac{\pi }{4}$ when $N_{11}(t)[=N_{22}(t)]$ is zero. At that point, the
Bloch vector $\overrightarrow{N}(t)=\{N_{03}(t),N_{30}(t),N_{33}(t)\}$ and
this state lies in one of the concordant subsets.

In both Markovian and non-Markovian regimes the interaction between the
qubits is kept the leading contribution to the total energy of the entire
system. As the noise strength is increased compatible to the interaction
term the discord decays much rapidly. \ We note some quantitative
differences in the Markovian and non-Makovian evolutions, but the evolution
categories do not change, since the origin of the categories is topological.

\textbf{Noise effect comparison on Ising- and XY-models.}

We note that the noise affects the quantum discord evolution in case of the
Ising model stronger than that in case of the anisotropic exchange
Hamiltonian (the $H^{XY}$ model). As $g_{z}$ is increased the discord
vanishes faster than that in the case of the $H^{XY}$ model.%

\begin{figure}[h]
\begin{center}
\vspace*{0.2cm} \includegraphics[width=9.5cm]{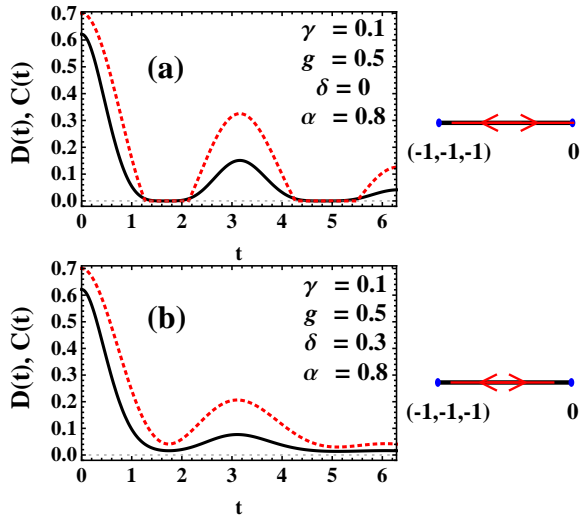}
\end{center}
\par
\vspace{-0.4cm}
\caption{(Color online) Joint evolution of discord (solid lines) and
entenglement (dashed lines) of a system subject to uncorrelated (a) and
correlated (b) telegraph noise. \ The switiching rate $\protect\gamma $ is
the same for the (a) and (b). \ The dynamics are non-Markovian with Ising
interaction and the intial state is a Werner state with $%
N_{xx}=N_{yy}=N_{zz}=-\protect\alpha =0.8.$ \ The sketches on the
right-hand-side represent the oscillatory trajectory along a line in the
direction (-1,-1,-1) in the Bloch-vector space. In (a) the trajectory hits
the origin in a finite time while in (b) it approaches the origin
asymptotically.}
\label{fig:2_noise_Un+corr_}
\end{figure}

\textbf{Noise correlation effect } 
We obtain the closed forms of the time dependence of all Bloch vector
components in case the qubits interact with each other through the Ising
spin exchange and with the two separate uncorrelated RTN sources of
different transition rates $\gamma_1=\gamma$, $\gamma_2 = \xi \gamma$ ($\xi
>0$). Full analytical solution for a general state can be obtained in
Appendix \ref{ap:uncorrelated_noise}. As shown in Appendix \ref%
{ap:uncorrelated_noise} the RTN noise on qubit 1 does not affect subclass (%
\ref{eq:Separable01_DQC1}), the RTN noise on qubit 2 does not affect
subclass (\ref{eq:Separable02_DQC1}), and the mutual qubit interaction does
not affect the \textbf{X}-type class [see Eq.~(\ref%
{eq:Bell_uncorrelated_gam12})]. This phenomenon is understood using
decoherence-free subspace theorem\cite{Lidar}. For $J=0$, all components in
each of the three separate above subclasses have similar time-dependent part
and only differ by their initial condition $N_{ij}(t=0)$. Enhancement of the
noise effect is seen in the \textbf{X}-type of class: e.g. for $\xi =1$ a
Bell-diagonal state at time $t$ has: 
\begin{eqnarray}  \label{eq:Uncorrelated_khac_Bell}
N_{ii} (t) &=& N_{ii} \,\, {\ e^{ - 2 \gamma t } \,\, G(R_0) }; \,\,\,\,
i=1, 2  \notag \\
N_{33}(t) &=& N_{33}
\end{eqnarray}
where 
\begin{eqnarray}  \label{eq:GR0}
G(R_0) &=& \frac {1} { {8R_0}^2 } [4 {g_z}^2 + ( 4 {R_0}^2 - {\gamma}^2 )
\cosh{\ (4 i R_0 t ) }  \notag \\
&& \,\,\,\,\,\,\,\,\,\,\,\,\,\, - 4 i \gamma R_0 \sinh{\ (4 i R_0 t )} ].
\end{eqnarray}
Fig.~\ref{fig:2_noise_Un+corr_}(a) shows the decoherence characteristic
using this model for a system initially prepared in a partially entangled
Werner state (\ref{eq:Werner_initial}) for $\alpha=0.8$. The quantum
entanglement exhibits category $\mathsf{O}$ i.e. the entanglement will
repeatedly disappear within some certain time period while the discord only
disappears at discrete time points.

The bottom sketch describes another non-Markovian process for the correlated
noise case where the trajectory never visits the origin and only approaches
this point as $t\rightarrow \infty .$ Both quantum entanglement and discord
never quite vanish in the $\mathsf{B}$-like category. 

\section{Conclusions}

The time evolution of quantum entanglement and quantum discord in 2-qubit
systems behave in fundamentally different ways. For the most part, this
difference comes from the different topologies of the zero sets: the set of
separable states and the set of concordant states, respectively. The set of
separable states is a convex 15-manifold. The set of concordant states is a
non-convex simply-connected when a certain set of zero measure has been
subtracted out. The generic time evolutions for the disappearance of
entanglement are of the $\mathsf{E}$ and $\mathsf{O}$ types, with $\mathsf{A}$ and $\mathsf{B}$ possible for symmetric
situations. The generic evolutions for discord disappearance are of $\mathsf{A}$ and $\mathsf{B}$
types, but discord disappearance depends on having the asymptotic limit
point lie on a set of low dimension: it is more rare than entanglement
disappearance, but it happens naturally in physical models, since, for
example, the completely mixed state is a common limit point. $\mathsf{E}$ and $\mathsf{O}$ types
of behavior are not allowed for discord. Furthermore, there are coexistence
rules for joint evolution. All these facts are summarized in the following
table. 
\begin{table}[tbp]
\begin{center}
\resizebox{7.5cm}{!}{
    \begin{tabular}{ | l | l | l | l | l | }
    \hline
    \text{Entanglement} &  A & E & B & O  \\ \hline
     \text{Discord} & A &  & B &  \\ \hline
      \text{Joint evolutions} & AA & EA, EB & BB & OB \\ \hline
    \end{tabular}}
\end{center}
\caption{Possible Routes for Disappearance of Entanglement and Discord in
Physical Systems. This table summarizes the possible ways that entanglement
and discord can disappear in physics models. Each joint evolution
corresponds to a different topology of two sets: the intersections of the
system trajectory with $\mathcal{S}$, the set of separable states, and with $%
\mathcal{C}$, the set of concordant states. We do not include pathological
trajectories with discontinuous derivatives or highly symmetrical models
whose trajectories are confined to low-dimensional submanifolds. \  }
\label{tab:Joint}
\end{table}
The time evolution of quantum entanglement and quantum discord in 2-qubit
systems behave in fundamentally different ways. \ For the most part, this
difference comes from the different topologies of the zero sets:\ the set of
separable states and the set of concordant states, respectively. \ The set
of separable states is a convex 15-manifold. \ The set of concordant states
is a non-convex simply-connected 9-manifold when certain sets of zero
measure have been subtracted out. \ The generic time evolutions for the
disappearance of the entanglement are of the $\mathsf{E}$ and $\mathsf{O}$ \
types, with $\mathsf{A}$ and $\mathsf{B}$ possible for highly symmetric
situations. \ The generic evolution of discord disappearance is of $\mathsf{A%
}$ and $\mathsf{B}$ types, but discord disappearance depends on having the
asymptotic limit point lie on a set of low dimension, so it is more rare
than entanglement disappearance. \ $\mathsf{E}$ and $\mathsf{O}$ type of
behavior do not occur. Roszak \textit{et al. }have computed the joint
evolution of entanglement and geometric discord in a model of two excitonic
quantum dot qubits dephased by noise from phonons\cite{Roszak}. They find
the expected phenomenon of incomplete disappearance of discord at long times
when the temperature is finite (and therefore the final state is not fully
mixed.) This case is not included in our analysis, though the generalization
is straightforward. \ In cases where the disappearance is complete, they
observe $\mathsf{EA}$ and $\mathsf{EB}$ behaviors for this model. Benedetti 
\textit{et al.} have done similar calculations for two qubits subjected to
classical noise\cite{Benedetti}. They observe $\mathsf{OB}$ and $\mathsf{BB}$
behavior, except when considering models that produce trajectories confined
to a low-dimensional manifold - in their case a mixture of Bell states.   \ 

Once the topology of the zero set is understood, the construction of
explicit models that display the various behaviors is relatively
straightforward. \ In particular one can show that qualitatively different
behaviors of entanglement and discord can be observed in the same system. \
This is true even if the evolution is unitary: with an Ising interaction one
can find oscillatory behavior of the discord even though the entanglement is
strictly zero at all times, for a judicious choice of the initial state. \
With a slightly more complicated Hamiltonian still with unitary time
development of the state, the coexistence of all reversible types of
oscillatory evolution for entanglement and discord can be obtained. \ For
example, $\mathsf{B}$\ or bouncing behavior of the discord is compatible
with the $\mathsf{O}$-type behavior of the concurrence in which the state is
separable for an infinite number of finite time intervals. \ For non-unitary
evolution, it is also found that all different kinds of evolutions of
discord and entanglement can coexist. \ We are able to produce coexistence
of $\mathsf{A}$ (half-life)\ type behavior of the discord with $\mathsf{E}$%
-type (entanglement sudden death), \ as well as $\mathsf{A}$ behavior for
both; coexistence of the various kinds of decaying oscillatory behavior is
also possible. \ 

It is not difficult to see how these evolutions correlate with various
topologies of the intersections of the state trajectory and the zero sets. \
For these considerations, the difference between different measures of
entanglement and the difference between geometric and quantum discord is not
material. \ For the question of frozen discord, however, the distinction
between geometric and quantum discord is essential. \ \ Frozen discord
occurs, by definition, when the trajectory is along a line of constant
discord. \ Since discord is unrelated to dynamics, this can occur (in the
absence of fine tuning)\ only when both the trajectory and the discord are
constrained by symmetry. \ Symmetry constraints typically lead to
straight-line trajectories. \ The surfaces of constant discord are not flat,
while the surfaces of constant geometric discord can be. \ Accordingly, we
only see frozen \textit{geometric} discord only in situations with high
symmetry, and frozen \textit{quantum} discord does not occur in natural
models.

\section{Acknowledgments}

This work is supported by DARPA QuEST program, Grant No. MSN118850. \ We
thank Hyungjun Lim for useful discussions.

\appendix

\section{The Bloch vector of a concordant state}

\label{a:obtainmu} 
In this section, we obtain the 15 components of the Bloch vector for a
concordant state. We define two projection operators: $\Pi _{k}\equiv |\Psi
_{k}\rangle \langle \Psi _{k}|=\frac{1}{2}(\sigma _{0}\pm \overrightarrow{m}%
\cdot \overrightarrow{\sigma })$ where $\overrightarrow{m}\equiv
(m_{1},m_{2},m_{3})=(\sin \theta \cos {\varphi },\sin {\theta }\sin {\varphi 
},\cos {\theta })$ is a Bloch unit vector. Qubit B is $\rho_k = \frac{1}{2}
( \sigma_0 \pm \overrightarrow{n}_k \cdot \overrightarrow{\sigma} ) $. %
We can rewrite Eq.~(\ref{eq:rhoc}) using these notations and the generalized
Bloch vector becomes: 
\begin{widetext}
\begin{eqnarray}\label{mu-ij-long}
N_{ij}&=&\frac{p}{2}\text{Tr}(\Pi_0\otimes \sigma_0)(\sigma_i\otimes \sigma_j)+\frac{1-p}{2}\text{Tr} (\Pi_1\otimes \sigma_0)(\sigma_i\otimes \sigma_j)+\\\nonumber
&&+\frac{p}{2}\sum_{k=1}^{3} {n_0}_{k} \text{Tr}(\Pi_0\otimes \sigma_k)(\sigma_i\otimes \sigma_j)+ \frac{1-p}{2}\sum_{k=1}^{3} {n_1}_{k}\text{Tr} (\Pi_1\otimes \sigma_k)(\sigma_i\otimes \sigma_j)\\\nonumber
&=& p \delta_{j,0} \text{Tr} \Pi_0\sigma_i +(1-p) \delta_{j,0} \text{Tr} \Pi_1\sigma_i\\\nonumber
&&+p(1-\delta_{j,0}){n_0}_j \text{Tr} \Pi_0\sigma_i + (1-p) (1-\delta_{j,0}) {n_1}_j \text{Tr} \Pi_1\sigma_i.\\\label{Eq:muif-finalE}
\Leftrightarrow N_{ij}&=&{p}[\delta_{j,0}+(1-\delta_{j,0}){n_0}_j]\text{Tr}\Pi_0\sigma_i + {(1-p)}[\delta_{j,0}+(1-\delta_{j,0}){n_1}_j]\text{Tr}\Pi_1\sigma_i.
\end{eqnarray}
\end{widetext}

These results can then be used to obtain the explicit forms for $N_{ij}$
given in (\ref{eq:mu1}), (\ref{eq:mu2}), and (\ref{eq:mu3}) of the main
text. 

\section{Concordant states on the coordinate axes}

\label{a:15axes} 
We wish to show that $\mathcal{C}$ contains the coordinate axes. \ 
\begin{eqnarray}
\rho &=&\frac{1}{4}(\sigma _{0}\otimes \sigma _{0}+\sum_{i}N_{0i}\sigma
_{0}\otimes \sigma _{i})  \notag  \label{eq:3axes} \\
&=&(\frac{1}{2}\sigma _{0})\otimes \frac{1}{2}(\sigma
_{0}+\sum_{i}N_{0i}\sigma _{i})  \notag \\
&=&\frac{1}{2}\Pi _{0}\otimes \rho _{0}+\frac{1}{2}\Pi _{1}\otimes \rho _{0}
\end{eqnarray}%
where $\Pi _{i}$; $i=0,1$ form an orthonormal basis in qubit A while $\rho
_{0}=\frac{1}{2}(\sigma _{0}+\sum_{i}N_{0i}\sigma _{i})$ is a general state
of qubit B given $N_{0i}\in \lbrack -1,1]$. This is always satisfied using
the condition of positivity for system qubit $\rho $. The last line in Eq. (%
\ref{eq:3axes}) is the necessary and sufficient condition for state $\rho $
to be concordant.


\section{Symmetries in the concordant subset}

\label{a:left_right} 
Generally, concordant states are asymmetric under exchange of the two
qubits. In this appendix, we show that a concordant state is symmetric under
some certain conditions. \newline
The left 0-discord state has the Bloch vector described as in Eqs.~(\ref%
{eq:mu1}), (\ref{eq:mu2}), (\ref{eq:mu3}) 
and \textquotedblleft left" geometric discord as in (\ref%
{eq:DG_analytical_form}) where 
\begin{eqnarray}
k_{\text{max}} &=&\sum_{i=1}^{3}{N_{i0}}^{2}+\sum_{i,j=1}^{3}{N_{ij}}^{2} 
\notag \\
&=&{\ |\underline{N}|}^{2}-({N_{01}}^{2}+{N_{02}}^{2}+{N_{03}}^{2}).
\end{eqnarray}%
%
%
A \textquotedblleft right" 0-discord state, instead, has: 
\begin{eqnarray}
N_{0i}^{^{\prime }} &=&(2p^{^{\prime }}-1)m_{i}^{^{\prime }}  \notag
\label{eq:Bloch_0_D_right} \\
N_{i0}^{^{\prime }} &=&p^{^{\prime }}n_{0i}^{^{\prime }}+(1-p^{^{\prime
}})n_{1i}^{^{\prime }}  \notag \\
N_{ij}^{^{\prime }} &=&m_{i}^{^{\prime }}[p^{^{\prime }}n_{0j}^{^{\prime
}}-(1-p^{^{\prime }})n_{1j}^{^{\prime }}]
\end{eqnarray}%
and its \textquotedblleft right" geometric discord is 
\begin{equation*}
{\mathcal{D}_{G}}^{R}=\frac{1}{4}[\text{Tr}(\overrightarrow{y}%
\overrightarrow{y}^{T})+\text{Tr}(\mathcal{T}^{^{\prime }}{\mathcal{T}%
^{^{\prime }}}^{T})-q_{\text{max}}]
\end{equation*}%
where $\overrightarrow{y}^{T}=(N_{10}^{^{\prime }},N_{20}^{^{\prime
}},N_{30}^{^{\prime }})$ and $q_{\text{max}}$ is the largest eigenvalues of $%
\overrightarrow{y}\overrightarrow{y}^{T}$ + $\mathcal{T}^{^{\prime }}{%
\mathcal{T}^{^{\prime }}}^{T}$. Generally, a left concordant state has ${%
\mathcal{D}_{G}}^{R}\neq 0$. Now, we discuss several cases which have $%
\mathcal{D}_{G}^{L}=0=\mathcal{D}_{G}^{R}$ i.e. symmetric concordant states.
If a (left) concordant state has $N_{0i}=\pm N_{i0}$, which is equivalent to
the condition $p\overrightarrow{n}_{0}+(1-p)\overrightarrow{n}_{1}=\pm (2p-1)%
\overrightarrow{m}$, then it is symmetric. Typical symmetric concordant
point is the origin which has $p=1/2$ and $\overrightarrow{n}_{i}=%
\overrightarrow{0}$. All pure concordant states satisfying two conditions $%
p(1-p)=0$ \& $|\overrightarrow{n}_{0}|\,|\overrightarrow{n}_{1}|=1$ are also
symmetric. 

\section{Linear independence of the tangent vectors}

\label{a:15tangents} 
We obtain the explicit form for the 9 tangent vectors to the manifold $%
\mathcal{C}$ as follows: 
\begin{widetext}
\begin{subequations}\label{eq:9tangents}
\begin{align}
\overrightarrow{t}_1=\frac{\partial{\overrightarrow{N}}}{\partial{\theta}}&=&\overrightarrow{0}  &	+	& (2p -1 )  \sum_{i=1}^3{\frac{\partial{m_i}}{\partial\theta}} \overrightarrow{e}_{i0} &+& \sum_{i,j=1}^3 [ {pn_{0j}-(1-p)n_{1j}} ] \frac{\partial m_i}{\partial\theta} \overrightarrow{e}_{ij}\label{eq:tangent01}\\
\overrightarrow{t}_2=\frac{\partial{\overrightarrow{N}}}{\partial{\varphi}}&=&\overrightarrow{0} &+	& ({2p-1})  \sum_{i=1}^3{\frac{\partial{m_i}}{\partial\varphi}} \overrightarrow{e}_{i0} &+& \sum_{i,j=1}^3 [ {pn_{0j}-(1-p)n_{1j}}]  \frac{\partial m_i}{\partial\varphi} \overrightarrow{e}_{ij}\label{eq:tangent02}\\
\overrightarrow{t}_3=\frac{\partial{\overrightarrow{N}}}{\partial{p}}& =  \sum_{i=1}^3 (n_{0i}-n_{1i}) &\overrightarrow{e}_{0i} &+& 	2 \sum_{i=1}^3 m_i \overrightarrow{e}_{i0} &+&  \sum_{i,j=1}^3 m_i(n_{0j}-n_{1j}) \overrightarrow{e}_{ij} \label{eq:tangent03} \\
\overrightarrow{t}_4=\frac{\partial{\overrightarrow{N}}}{\partial{n_{01}}}&= p \sum_{i=1}^3 \delta_{i,1} &\overrightarrow{e}_{0i} &+& \overrightarrow{0} &+& p \sum_{i=1}^3 m_i\delta_{j,1} \overrightarrow{e}_{ij} \label{eq:tangent04}\\
\overrightarrow{t}_5=\frac{\partial{\overrightarrow{N}}}{\partial{n_{02}}}&= p \sum_{i=1}^3 \delta_{i,2} &\overrightarrow{e}_{0i} &+& \overrightarrow{0} &+& p  \sum_{i=1}^3 m_i\delta_{j,2} \overrightarrow{e}_{ij}\label{eq:tangent05} \\
\overrightarrow{t}_6=\frac{\partial{\overrightarrow{N}}}{\partial{n_{03}}}&= p \sum_{i=1}^3 \delta_{i,3} & \overrightarrow{e}_{0i} &+& \overrightarrow{0} &+& p \sum_{i=1}^3 m_i\delta_{j,3} \overrightarrow{e}_{ij} \label{eq:tangent06}\\
\overrightarrow{t}_7=\frac{\partial{\overrightarrow{N}}}{\partial{n_{11}}}&= (1-p) \sum_{i=1}^3 \delta_{i,1} &\overrightarrow{e}_{0i} &+& \overrightarrow{0} &-& (1-p) \sum_{i=1}^3 m_i\delta_{j,1} \overrightarrow{e}_{ij} \label{eq:tangent07} \\
\overrightarrow{t}_8=\frac{\partial{\overrightarrow{N}}}{\partial{n_{12}}}&= (1-p) \sum_{i=1}^3 \delta_{i,2} & \overrightarrow{e}_{0i} &+& \overrightarrow{0} &-& (1-p) \sum_{i=1}^3 m_i\delta_{j,2} \overrightarrow{e}_{ij} \label{eq:tangent08} \\
\overrightarrow{t}_9=\frac{\partial{\overrightarrow{N}}}{\partial{n_{13}}}&= (1-p) \sum_{i=1}^3 \delta_{i,3}  &\overrightarrow{e}_{0i} &+& \overrightarrow{0} &-& (1-p) \sum_{i=1}^3 m_i\delta_{j,3} \overrightarrow{e}_{ij} \label{eq:tangent09}.
\end{align}
\end{subequations}
\end{widetext}
The tensor matrix has explicit form as: %
\begin{equation*}
g= \left( 
\begin{array}{ccccccccc}
g_{11} & 0 & 0 & 0 & 0 & 0 & 0 & 0 & 0 \\ 
0 & g_{22} & 0 & 0 & 0 & 0 & 0 & 0 & 0 \\ 
0 & 0 & g_{33} & g_{34} & g_{35} & g_{36} & 0 & 0 & 0 \\ 
0 & 0 & g_{43} & g_{44} & 0 & 0 & 0 & 0 & 0 \\ 
0 & 0 & g_{53} & 0 & g_{55} & 0 & 0 & 0 & 0 \\ 
0 & 0 & g_{63} & 0 & 0 & g_{66} & 0 & 0 & 0 \\ 
0 & 0 & 0 & 0 & 0 & 0 & g_{77} & 0 & 0 \\ 
0 & 0 & 0 & 0 & 0 & 0 & 0 & g_{88} & 0 \\ 
0 & 0 & 0 & 0 & 0 & 0 & 0 & 0 & g_{99} \\ 
&  &  &  &  &  &  &  & 
\end{array}
\right)
\end{equation*}
with $g_{11} = (1-2p)^2 + \sum_{i=1}^{3} {\ [ pn_{0i} - (1-p) n_{1i}]^2 }; $ 
$g_{22} = \{(1-2p)^2 + \sum_{i=1}^{3} {\ [ pn_{0i} - (1-p) n_{1i}]^2 } \}
\sin^2{\theta};$ $g_{33} = 2 [ 2+ \sum_{i=1}^{3} {(n_{0i} -n_{1i})^2} ];$ $%
g_{44}=g_{55}=g_{66}= 2p^2;$ $g_{77}=g_{88}=g_{99}=2(1-p)^2;$ $%
g_{34}=g_{43}=2p(n_{01}-n_{11});$ $g_{35}=g_{53}=2p(n_{02}-n_{12});$ $%
g_{36}=g_{63}=2p(n_{03}-n_{13})$.

\section{Quantum discord of the state in Eq.~(\ref{eq:DQC1_unitary})}

\label{ap:discord_calculation} 
The corresponding density matrix $\rho(t) = \frac{1}{4} ( \sigma_0 \otimes
\sigma_0 + S_3 \sigma_{1} \otimes \sigma_3 + C_3 \sigma_2 \otimes \sigma_0 ) 
$ 
has 4 eigenvalues of $\{\frac{1}{2},\frac{1}{2},0,0\}$ and $S[\rho(t)]=1$. 
The two subsystems are: $\rho_A(t) = \frac{1}{2} \left( 
\begin{array}{cc}
1 & -i C_3 \\ 
i C_3 & 1%
\end{array}
\right) $ with two eigenvalues $\lambda_A=\frac{1}{2} (1 \pm C_3)$; and $%
\rho_B(t)=\sigma_0/2$. 

The quantum mutual information is: 
\begin{eqnarray}
\mathcal{I} &=&S[\rho _{A}(t)]+S[\rho _{B}(t)]-S[\rho (t)]  \notag
\label{eq:q_mutual} \\
&=&S[\rho _{A}(t)]  \notag \\
&=&1-\frac{1}{2}[(1-C_{3})\log (1-C_{3})  \notag \\
&&+(1+C_{3})\log (1+C_{3})].
\end{eqnarray}
The classical mutual information is 
\begin{eqnarray}  \label{eq:c_mutual}
\mathcal{J^{\text{class}}}&=&S[\rho_{B}(t)] - \min_{\{A_k\}} S[\rho(t)|A_k] 
\notag \\
&=& 1 - \min_{\{A_k\}} S[\rho(t)|A_k]
\end{eqnarray}

where $\{A_k=V \Pi_k V^{\dagger} ; \,\,k=1,\,2 \} $ defines a set of
measurement on subsystem A: 
$V =t \sigma_0 + \overrightarrow{v} \cdot \overrightarrow{\sigma} $ 
where $t^2 + {v_1}^2 + {v_2}^2 + {v_3}^2 = 1$ and $\{ \Pi_k \}$ is some
local orthogonal basis. Without loss of generality, we choose $\Pi_1= \left( 
\begin{array}{cc}
1 & 0 \\ 
0 & 0%
\end{array}
\right) $ and $\Pi_2= \left( 
\begin{array}{cc}
0 & 0 \\ 
0 & 1%
\end{array}
\right).$ 
Some useful expressions are used: 
$\Pi_1 \sigma_i \Pi_1 =\delta_{i,3} \Pi_1; \,\,\,\,\,\,\,\, \Pi_2 \sigma_i
\Pi_2 =-\delta_{i,3} \Pi_2 $, 
\begin{eqnarray}
z_1=\text{ Tr} {V^{\dagger} \sigma_{1} V \sigma_{3} }&=&2(-t v_2 + v_1v_3),
\\
z_2=\text{ Tr} {V^{\dagger} \sigma_{2} V \sigma_{3} }&=& 2(t v_1 + v_2 v_3 ),
\\
z_3=\text{ Tr} {V^{\dagger} \sigma_{3} V \sigma_{3} }&=&t^2+{v_3}^2 - {v_1}%
^2 -{v_2}^2. \\
\notag
\end{eqnarray}
After measurement $\{A_k\}$ the system is sent to 
${\rho_k}(t)=\frac{A_k \rho(t) A_k}{\text{Tr} ( A_k \rho(t) A_k ) }$. 
Set $p_k=\text{Tr} ( A_k \rho(t) A_k ) $ with 
$p_1=\frac{1+X}{2}, p_2=\frac{1-X}{2}$ 
where $X=2 C_3 (tv_1 + v_2 v_3) = z_2 C_3$, one obtains: 
\begin{widetext}
\begin{eqnarray}\label{eq:rho1_U_t}
p_1\rho_1(t) &=&  A_1 \rho(t) A_1\nonumber\\
&=& \frac{1}{4} (V \otimes \sigma_0) (\Pi_1 \otimes \sigma_0 ) \nonumber\\
&&\hspace{1cm}   (V^{\dagger}  \otimes \sigma_0 )    (\sigma_0 \otimes \sigma_0  + C_3 \sigma_2 \otimes \sigma_0 + S_3 \sigma_1 \otimes \sigma_3     )
 (V \otimes \sigma_0 ) \nonumber\\
&&\hspace{5cm} ( \Pi_1  \otimes \sigma_0 ) ( V^{\dagger} \otimes \sigma_0 )\nonumber\\
p_1 \rho_1(t) &=&\frac{1}{4} (V \Pi_1 V^{\dagger}) \otimes [ (1 + X) \sigma_0 + Y \sigma_3  ]
\end{eqnarray}
\end{widetext}
with X defined above and $Y=z_1S_3$.

Similarly, $\rho_2(t)$ is obtained as: 
\begin{eqnarray}  \label{eq:rho2_U_t}
p_2 \rho_2 (t) &=& (V \Pi_1 V^{\dagger}) \otimes [ (1 - X) \sigma_0 - Y
\sigma_3 ].
\end{eqnarray}
Now, eigenvalues of $\rho_1(t)$ are $\frac{1}{4p_1} (1+X+Y); \frac{1}{4p_1}
(1+X-Y)\,\, $ and of $\rho_2(t)$ are $\frac{1}{4p_2} (1-X+Y); \frac{1}{4p_2}
(1-X-Y)\,\, $. One can obtain the conditional entropy as: 
\begin{widetext}
\begin{eqnarray}  \label{eq:condt_S}
S[\rho(t)|A_k] &=&  p_1 S[\rho_1(t)] + p_2 S[\rho_2(t)] \nonumber\\
                    &=& 1 - \frac{1}{4} [ (1+X+Y)\log(1+X+Y)  +  (1+X-Y)\log(1+X-Y)  \nonumber \\ 
                    &&+  (1-X+Y)\log(1-X+Y)  + (1-X-Y)\log(1-X-Y)  \nonumber\\
                    && - 2 (1+X) \log(1+X) - 2 (1-X) \log(1-X)      ].
\end{eqnarray}
\end{widetext}

Note that $X=2 C_3 (tv_1 + v_2 v_3) \leq |C_3 | $; $Y=2 S_3 (-tv_2+v_1 v_3)
\leq |S_3|$. (\ref{eq:condt_S}) has identical minima at $X=0$ \& $Y=\pm S_3$%
. As a result, 
\begin{eqnarray}  \label{eq:condt_S_min}
\min_{\{ A_k\}} S [\rho(t)|A_k] &=& 1 - \frac{1}{2} [ (1-|S_3|)\log(1-|S_3|)
\notag \\
&& + (1+|S_3|)\log(1+|S_3|) ]
\end{eqnarray}
Now substitute (\ref{eq:condt_S_min}) into (\ref{eq:c_mutual}) and subtract (%
\ref{eq:c_mutual}) from (\ref{eq:q_mutual}) the quantum discord is obtained
as in the main text. 

\section{Quantum discord of the state in Eq. (\ref{eq:Bloch_RTN_ZZ})}

\label{ap:DQC1_RTN} 
Note that 
\begin{eqnarray}
| \overrightarrow{N} (t) | &=& \sqrt { {N_{20}}^2 (t) + {N_{13}}^2 (t) } 
\notag \\
&=& e^{- \gamma t} | \cosh{(2 i R_0 t)} - i \frac{ \gamma} {2R_0} \sinh{\ (2
i R_0 t )} |.  \notag \\
\end{eqnarray}
%
%
Entropy of the system (\ref{eq:Bloch_RTN_ZZ}) is 
\begin{eqnarray}
S_{\rho} &=& - ( \lambda_1 \log{\lambda_1} + \lambda_2 \log{\lambda_2} +
\lambda_3 \log{\lambda_3}  \notag \\
&& + \lambda_4 \log{\lambda_4} )
\end{eqnarray}
where 
\begin{eqnarray}
\lambda_{1,2} &=& \frac{1}{4} [ 1 + \sqrt{{N_{20}}^2 (t) + {N_{13}}^2 (t) } ]
\notag \\
\lambda_{3,4} &=& \frac{1}{4} [ 1 - \sqrt{{N_{20}}^2 (t) + {N_{13}}^2 (t) }
].
\end{eqnarray}
Entropy of system A is 
\begin{eqnarray}
S_{\rho_A} &=& 1 - \frac{1}{2} \{ [1+N_{20} (t)] \log{\ [1+N_{20} (t) ]} 
\notag \\
&+ & [1-N_{20} (t)] \log{\ [1-N_{20} (t)]} \}
\end{eqnarray}
and the classical mutual information after optimization is: 
\begin{eqnarray}
S_{\text{class.}} &=& 1 - \frac{1}{2} \{ [1+N_{13} (t)] \log{\ [1+N_{13} (t)
]}  \notag \\
&& + [1-N_{13} (t)] \log{\ [1-N_{13} (t)]} \}.
\end{eqnarray}
The time evolution of the quantum discord of the above system: 
$\mathcal{D} (t) = S_{\rho_A } - S_{\rho} + S_{\text{class.}}. $ 

\section{Quantum discord of the state in Eq. (\ref{eq:NtBell_ZZ_RTN})%
}

\label{ap:discord_Werner_Ising_RTN} 
\begin{eqnarray}
\mathcal{D} (t) &=& 2 - \frac{1+\alpha}{2} \log{(1+\alpha)} - \frac{1-\alpha%
}{2} \log{(1-\alpha)}  \notag \\
&& - S_{\rho}
\end{eqnarray}
where 
$
S_{\rho} = - \lambda_0 \log{\lambda_0} - \lambda_1 \log{\lambda_1} -
\lambda_2 \log{\lambda_2} - \lambda_3 \log{\lambda_3} $ 
with $\lambda_0=\lambda_1=\frac{1-\alpha}{4}$; $\lambda_{2,3}=\frac{1+\alpha
\pm 2 \sqrt{ {N_{11}}^2 (t) + {N_{12}}^2 (t) } }{4}$.

\section{Quasi-Hamiltonian $H_q$ of the XY-model}

\label{ap:XY_Hq} 
This section is for further discussions of solving the XY-model in a
non-unitary evolution.\newline
In a general XY-model in interaction with a single RTN fluctuator ($B_z=0$), 
$H_q$ has 15 different eigenvalues: $-2 i \gamma, -i \gamma, -2 R_0,
-i\gamma \pm 2 R_0, -i \gamma \pm W_1, - i \gamma \pm W_2, \omega_1,
\omega_2, \omega_3, \Omega_1, \Omega_2, \Omega_3$ where $W_{1,2}=\sqrt{ 4 {%
J_{yx}}^2 + 2 {g_z}^2 -{\gamma}^2 \pm \sqrt{ {g_z}^4 + 4 {J_{yx}}^2 {g_z}^2
- 4 {J_{yx}}^2 \gamma^2 } }$; $\omega_i$ and $\Omega_i$, respectively, are
roots of polynomial $-32 J_{yx}^2 -4 (4 J_{yx}^2 + {g_z}^2) \omega + 2 i
\gamma {\omega}^2 + {\omega}^3$ and $- 8 i {g_z}^2 \gamma - 4 (4 J_{yx}^2 + {%
g_z}^2 + \gamma^2) {\Omega} + 4 i \gamma {\Omega}^2 + {\Omega}^3 $. 

\section{Analytical solution of two uncorrelated RTN fluctuators with
different transition rates $\protect\gamma, $ $\protect\xi \protect\gamma$}

\label{ap:uncorrelated_noise} 
We obtain exact solutions for the Ising models in interaction with two
different uncorrelated noise sources.\newline
For the components constructing a general off-\textbf{X}-type of class [see
the left-hand-side (LHS) of Eqs. (\ref{eq:Separable01_DQC1}) and (\ref%
{eq:Separable02_DQC1})], the solution yields by replacing $\gamma$ as in the
single fluctuator case [see Eq.~(\ref{eq:Bloch_RTN_ZZ})] by $\gamma_1, $ $%
\gamma_2$ (respectively, equals $\gamma$, $\xi \, \gamma$) for the
corresponding subclass. It is due to the fact that the LHS components in Eq.
(\ref{eq:Separable01_DQC1}) commute with the term $\sigma_0 \otimes \sigma_3$
and the LHS components in Eq. (\ref{eq:Separable02_DQC1}) commute with the
term $\sigma_3 \otimes \sigma_0$. As a result, each subclass is not affected
by the noise from the other qubit. More generally, if a Hamiltonian consists
of $k$ commute terms then the combined solution is $N_{ij}(t) = \sum_{i,j} {%
\ N_{ij} \Pi_{m=1}^{k} f^{m}_{ij} (t)}$ where $\sum_{i,j} N_{ij}
f_{ij}^{m}(t)$ is the solution for the $m$-th term. Back to the above case,
one has: 
\begin{eqnarray}
N_{01}(t) &=& (N_{01} C_3 + N_{32} S_3 ) e^{- \xi \gamma t} F(R_2)  \notag \\
N_{32}(t) &=& (N_{32} C_3 - N_{01} S_3 ) e^{- \xi \gamma t} F(R_2)  \notag \\
N_{02}(t) &=& (N_{30} C_3 - N_{31} S_3 ) e^{- \xi \gamma t} F(R_2)  \notag \\
N_{31}(t) &=& (N_{31} C_3 + N_{02} S_3 ) e^{- \xi \gamma t} F(R_2);
\end{eqnarray}

\begin{eqnarray}
N_{10}(t) &=& (N_{10} C_3 + N_{23} S_3 ) e^{- \gamma t} F(R_1)  \notag \\
N_{23}(t) &=& (N_{23} C_3 - N_{10} S_3 ) e^{- \gamma t} F(R_1)  \notag \\
N_{20}(t) &=& (N_{03} C_3 - N_{13} S_3 ) e^{- \gamma t} F(R_1)  \notag \\
N_{13}(t) &=& (N_{13} C_3 + N_{20} S_3 ) e^{- \gamma t} F(R_1)
\end{eqnarray}
where $R_{1,2}=\sqrt{ 4 {g_z}^2 - {\gamma_{1,2}}^2 }/2.$

The enhanced noise effect will be seen in the other class - the \textbf{X}%
-type: 
\begin{eqnarray}  \label{eq:Bell_uncorrelated_gam12}
N_{ij}(t) &=& N_{ij} e^{- \gamma (1+ \xi) t} H(R_0,\xi), \,\,\,\,\, i, j =1
\div 3  \notag \\
N_{33}(t)&=& N_{33}  \notag \\
\end{eqnarray}
where 
\begin{eqnarray}
H(R_0,\xi) &=& \frac{1}{4 R_0 X_0} [ 2 R_0 \cosh{(2 i R_0 t)} - i \gamma
\sinh{\ (2 i R_0 t )} ]  \notag \\
&&\,\,\,\,\,\, [ 2 X_0 \cosh{\ (2 i X_0 t )} - i \xi \gamma \sinh{\ (2 i X_0
t )} ].  \notag \\
\end{eqnarray}
Here, $X_0=\sqrt{ 4 {R_0}^2 - \gamma^2 ({\xi}^2 -1) } /2$. If $\xi=1$ (i.e. $%
\gamma_1=\gamma_2$) then $X_0$=$R_0$ and $H(R_0,\xi) \equiv G(R_0)$ as
defined in (\ref{eq:GR0}).


\begin{thebibliography}{99}
\bibitem{Schrodinger} E. Schr\"{o}dinger, Proc. Cambridge Philos. Soc. 
\textbf{31}, 555 (1935).

\bibitem{Vedral} L. Henderson and V. Vedral, J. Phys. A \textbf{34}, 6899
(2001).

\bibitem{Zurek} H. Ollivier and W. H. Zurek, Phys. Rev. Lett. \textbf{88},
017901 (2001).

\bibitem{Laflamme} E. Knill and R. Laflamme, Phys. Rev. Lett. \textbf{81},
5672 (1998).

\bibitem{Datta} A. Datta, S. T. Flammia, and C. M. Caves, Phys. Rev. A 
\textbf{72}, 042316 (2005).

\bibitem{Datta2} A. Datta, A. Shaji, and C. M. Caves, Phys. Rev. Lett. 
\textbf{100}, 050502 (2008).

\bibitem{Bennett} C.H. Bennett and S.J. Wiesner, Phys. Rev. Lett. \textbf{69}%
, 2881 (1992).

\bibitem{Modi} K. Modi, A. Brodutch, H. Cable, T. Paterek, and V. Vedral,
Rev. Mod. Phys. \textbf{84}, 1655 (2012).

\bibitem{Yu} T. Yu and J. H. Eberly, Science \textbf{323}, 598 (2009).

\bibitem{Werlang_2009} T. Werlang, S. Souza, F. F. Fanchini, and C. J.
Villas Boas, Phys. Rev. A \textbf{80}, 024103 (2009).

\bibitem{Maniscalco} L. Mazzola, J. Piilo, and S. Maniscalco, Phys. Rev.
Lett. \textbf{104}, 200401 (2010).

\bibitem{Ferraro2010} A. Ferraro, L. Aolita, D. Cavalcanti, F. M.
Cucchietti, and A. Acin\'{\i}n, Phys. Rev. A \textbf{81}, 052318 (2010).

\bibitem{Fanchini} F. F. Fanchini, T. Werlang, C. A. Brasil, L. G. E.
Arruda, and A. O. Caldeira, Phys. Rev. A \textbf{81}, 052107 (2010).

\bibitem{Fanchinivolume} F. F. Fanchini, L. K. Castelano, and A. O.
Caldeira, New J. Phys. \textbf{12}, 073009 (2010).

\bibitem{Lang2011} M. D. Lang, C. C. Caves, and A. Shaji, Int. J. Quantum
Inform. \textbf{9}, 1553 (2011).

\bibitem{Mazzola} L. Mazzola, J. Piilo, and S. Maniscalco, Int. J. Quantum.
Inform. \textbf{9}, 981 (2011).

\bibitem{Bellomo2012} B. Bellomo, R. Lo Franco, and G. Compagno, Phys. Rev.
A \textbf{86}, 012312 (2012).

\bibitem{Pal} A. K. Pal and I. Bose, Eur. Phys. J. B \textbf{85}, 277 (2012).

\bibitem{Zhou1} D. Zhou and R. Joynt, Quantum Inf. Process \textbf{11}, 571
(2012).

\bibitem{Zhou2} D. Zhou, G. W. Chern, J. Fei, and R. Joynt, Int. J. Mod.
Phys. B \textbf{26}, 1250054 (2012).

\bibitem{Caves} M. D. Lang and C. M. Caves, Phys. Rev. Lett. \textbf{105},
150501 (2010).

\bibitem{Roszak} K. Roszak, P. Mazurek, and P. Horodecki, Phys. Rev. A 
\textbf{87}, 062308 (2013).

\bibitem{Benedetti} C. Benedetti, F. Buscemi, P. Bordone, and M. Paris,
Phys. Rev. A 87, 052328 (2013).

\bibitem{Luo} S. Luo, Phys. Rev. A \textbf{77}, 042303 (2008).

\bibitem{Ali} M. Ali, A. R. P. Rau, and G. Alber, Phys. Rev. A \textbf{81},
042105 (2010).

\bibitem{Girolami2011} D. Girolami and G. Adesso, Phys. Rev. A \textbf{83},
052108 (2011); Phys. Rev. A \textbf{84}, 052110 (2011).

\bibitem{Dakic} B. Daki\'{c}, V. Vedral, and C. Brukner, Phys. Rev. Lett. 
\textbf{105}, 190502 (2010).

\bibitem{Girolami} D. Girolami and G. Adesso, Phys. Rev. Lettt. \textbf{108}%
, 150403 (2012).

\bibitem{Jakobczyk} L. Jak\'{o}bczyk and M. Siennicki, Phys. Lett. A \textbf{%
286}, 383 (2001).

\bibitem{LuoFu} S. Luo and S. Fu, Phys. Rev. A \textbf{82}, 034302 (2010).

\bibitem{Lim} H. Lim and R. Joynt (in preparation).

\bibitem{Bennett2} C. H. Bennett, D. P. DiVincenzo, J. Smolin, and W. K.
Wootters, Phys. Rev. A \textbf{54}, 3824 (1996).

\bibitem{Wootters} W. K. Wootters, Phys. Rev. Lett. \textbf{80}, 2245 (1998).

\bibitem{Werner} R. F. Werner, Phys. Rev. A \textbf{40}, 4277 (1989).

\bibitem{Cheng} B. Cheng, Q. H. Wang, and R. Joynt, Phys. Rev. A \textbf{78}%
, 022313 (2008).

\bibitem{Bob} R. Joynt, D. Zhou, and Q. H. Wang, Int. J. Mod. B \textbf{25},
2115 (2011); D. Zhou and R. Joynt, Phys. Rev. A \textbf{81}, 010103 (2010).

\bibitem{Lidar} D. A. Lidar, I. L. Chuang, and K. B. Whaley, Phys. Rev.
Lett. \textbf{81}, 2594 (1998).
\end{thebibliography}
\end{document}